\documentclass[12pt,preprint]{aastex}
\usepackage{lscape}
\def\CIVdblt{{\rm C}\kern 0.1em{\sc iv}~$\lambda\lambda 1548, 1550$}
\def\MgIIdblt{{\rm Mg}\kern 0.1em{\sc ii}~$\lambda\lambda 2796, 2803$}
\def\NVdblt{{\rm N}\kern 0.1em{\sc v}~$\lambda\lambda 1238, 1242$}  
\def\OVIdblt{{\rm O}\kern 0.1em{\sc vi}~$\lambda\lambda 1031, 1037$} 
\def\SiIVdblt{{\rm Si}\kern 0.1em{\sc iv}~$\lambda\lambda1394, 1403$}  

\def\AlII{\hbox{{\rm Al}\kern 0.1em{\sc ii}}}
\def\AlIII{{\hbox{\rm Al}\kern 0.1em{\sc iii}}}
\def\CaI{\hbox{{\rm Ca}\kern 0.1em{\sc i}}}
\def\CaII{\hbox{{\rm Ca}\kern 0.1em{\sc ii}}}
\def\CrII{\hbox{{\rm Cr}\kern 0.1em{\sc ii}}}
\def\CII{\hbox{{\rm C}\kern 0.1em{\sc ii}}}
\def\CIII{\hbox{{\rm C}\kern 0.1em{\sc iii}}}
\def\CIV{\hbox{{\rm C}\kern 0.1em{\sc iv}}}
\def\CV{\hbox{{\rm C}\kern 0.1em{\sc v}}}
\def\HI{\hbox{{\rm H}\kern 0.1em{\sc i}}}
\def\HII{\hbox{{\rm H}\kern 0.1em{\sc ii}}}
\def\Lya{\hbox{{\rm Ly}\kern 0.1em$\alpha$}}
\def\Lyb{\hbox{{\rm Ly}\kern 0.1em$\beta$}}
\def\Lyg{\hbox{{\rm Ly}\kern 0.1em$\gamma$}}
\def\Lyfive{\hbox{{\rm Ly}\kern 0.1em$5$}}
\def\Lysix{\hbox{{\rm Ly}\kern 0.1em$6$}}
\def\Lyseven{\hbox{{\rm Ly}\kern 0.1em$7$}}
\def\Lyeight{\hbox{{\rm Ly}\kern 0.1em$8$}}
\def\Lynine{\hbox{{\rm Ly}\kern 0.1em$9$}}
\def\Lyten{\hbox{{\rm Ly}\kern 0.1em$10$}}
\def\HeI{\hbox{{\rm He}\kern 0.1em{\sc i}}}
\def\HeII{\hbox{{\rm He}\kern 0.1em{\sc ii}}}
\def\FeI{\hbox{{\rm Fe}\kern 0.1em{\sc i}}}
\def\FeII{\hbox{{\rm Fe}\kern 0.1em{\sc ii}}}
\def\FeIII{\hbox{{\rm Fe}\kern 0.1em{\sc iii}}}
\def\MnII{\hbox{{\rm Mn}\kern 0.1em{\sc ii}}}
\def\MgI{\hbox{{\rm Mg}\kern 0.1em{\sc i}}}
\def\MgII{\hbox{{\rm Mg}\kern 0.1em{\sc ii}}}
\def\MgIII{\hbox{{\rm Mg}\kern 0.1em{\sc iii}}}
\def\MgIV{\hbox{{\rm Mg}\kern 0.1em{\sc iv}}}
\def\NaI{\hbox{{\rm Na}\kern 0.1em{\sc i}}}
\def\NV{\hbox{{\rm N}\kern 0.1em{\sc v}}}
\def\NII{\hbox{{\rm N}\kern 0.1em{\sc ii}}}
\def\NIII{\hbox{{\rm N}\kern 0.1em{\sc iii}}}
\def\OVI{\hbox{{\rm O}\kern 0.1em{\sc vi}}}
\def\OII{\hbox{[{\rm O}\kern 0.1em{\sc ii}]}}
\def\SiII{\hbox{{\rm Si}\kern 0.1em{\sc ii}}}
\def\SiIII{\hbox{{\rm Si}\kern 0.1em{\sc iii}}}
\def\SiIV{\hbox{{\rm Si}\kern 0.1em{\sc iv}}}
\def\SII{\hbox{{\rm S}\kern 0.1em{\sc ii}}}
\def\SIII{\hbox{{\rm S}\kern 0.1em{\sc iii}}}
\def\SIV{\hbox{{\rm S}\kern 0.1em{\sc iv}}}
\def\TiII{\hbox{{\rm Ti}\kern 0.1em{\sc ii}}}
\def\ZnII{\hbox{{\rm Zn}\kern 0.1em{\sc ii}}}
\newcommand{\kms}{\hbox{km~s$^{-1}$}}

\def\kms{\hbox{km~s$^{-1}$}}       
\def\cm2{\hbox{cm$^{-2}$}}

\def\etal{et~al.\ }

\def\Mgn{\hbox{{\rm Mg}}}
\def\Mgi{\hbox{{\rm Mg}$^+$}}
\def\Mgl{\hbox{$^{24}${\rm Mg}}}
\def\Mg24i{\hbox{$^{24}${\rm Mg}$^+$}}
\def\Hn{\hbox{{\rm H}}}

\begin{document}
\title{The Advantage of Increased Resolution in the Study of Quasar Absorption Systems}
\author{Anand~Narayanan\altaffilmark{1},Toru~Misawa\altaffilmark{1}, Jane~C.~Charlton\altaffilmark{1}, and Rajib~Ganguly\altaffilmark{2}}

\altaffiltext{1}{Department of Astronomy and Astrophysics, The Pennsylvania State University, University Park, PA 16802, {\it anand, misawa, charlton@astro.psu.edu}}
\altaffiltext{2}{Department of Physics and Astronomy, 1000 East University Ave, University of Wyoming (Dept 3905), Laramie, WY, 82071, ganguly@uwyo.edu}

\begin{abstract}
We compare a new $R=120,000$ spectrum of PG~$1634+706$ ($z_{QSO}=1.337,m_V=14.9$) obtained with the HDS instrument on Subaru to a $R=45,000$ spectrum obtained previously with HIRES/Keck.
In the strong {\MgII} system at $z=0.9902$ and the multiple cloud, weak
{\MgII} system at $z=1.0414$, we find that at the higher resolution,
additional components are resolved in a blended profile.  We find
that two single-cloud weak {\MgII} absorbers were already resolved
at $R=45,000$, to have $b=2$-$4$~{\kms}.  The narrowest line that
we measure in the $R=120,000$ spectrum is a component of the Galactic
{\NaI} absorption, with $b=0.90\pm0.20$~{\kms}.  We discuss expectations
of similarly narrow lines in various applications, including studies
of DLAs, the {\MgI} phases of strong {\MgII} absorbers, and high
velocity clouds. By applying Voigt profile fitting to synthetic lines,
we compare the consistency with which line profile parameters can be 
accurately recovered at $R=45,000$ and $R=120,000$. We estimate the
improvement gained from superhigh resolution in resolving narrowly
separated velocity components in absorption profiles. We also explore
the influence of isotope line shifts and hyperfine splitting in 
measurements of line profile parameters, and the spectral resolution
needed to identify these effects. Super high resolution spectra of quasars, which
will be routinely possible with 20-meter class telescopes, will lead
to greater sensitivity for absorption line surveys, and to determination
of more accurate physical conditions for cold phases of gas in
various environments.
\end{abstract}

\section{INTRODUCTION}
\label{sec:1}

With the advent of high resolution spectroscopy on 8-meter class
telescopes in the early 1990's came a revelation in the study of
quasar absorption line systems.  Previous to this time, several large
surveys were conducted with a typical resolution of $R =
\lambda/\Delta \lambda = 1000$--$3000$, corresponding to
$100$--$300$~{\kms}.  These surveys tabulated the redshift path
densities of {\MgII}, {\CIV}, and {\Lya} lines over the redshift range
accessible from optical telescopes \citep*{lanzetta87, sbs88, caulet}.
However, most individual systems were not resolved, so that the kinematics of the
different metal lines could not be compared system by system.  It is
not surprising that an increase of resolution to $R=45,000$
($\sim7$~{\kms}) provided detailed views of the multiphase structure
of gas inside and around galaxies, since parcels of gas within the
different components of galaxies are often moving at several to tens
of {\kms} relative to one another.  The increase of resolution from
thousands to tens of thousands also brought a dramatic increase in the
sensitivity of surveys.  This allowed measurements of the metallicity
in the most diffuse gas sampled by weak {\Lya} forest lines
\citep[e.g., ][]{cowie95}, and for deeper surveys of metal lines as well
\citep[e.g., ][]{weak1,toru02}.

Presently, hundreds of quasar spectra have been obtained with
$R\sim45,000$ using the High Resolution Echelle Spectrograph on Keck I \citep[HIRES, ][]{hires}, the High Dispersion Spectrograph on Subaru \citep[HDS, ][]{hds} and the Ultraviolet-Visual Echelle Spectrograph on the VLT \citep[UVES, ][]{uves}.  There is undoubtedly much
to be gained from detailed studies of the absorption lines detected in
these databases.  However, in this paper we look toward the future and
ask what advances we should expect in studies of quasar absorption
lines with an increase in resolution from the presently attainable
value of $7$~{\kms} to $0.5$-$2$~{\kms}, which is within the reach of
20-meter class telescopes for hundreds of quasars.  This follows a
study by \citet{tappeblack} of the quasar PKS~$2145+067$ at $R\sim100,000$
with UVES on the VLT UT2 telescope.  That study focused on the $z=0.79089$
strong {\MgII} absorber, and measured $b$ parameters in the range
$1$--$7$~{\kms}, showing considerably more structure than previous
spectra \citep{cwcvogt01} with $R\sim45,000$.  More recently,
\citet{chand06} considered the effect of observing at $R>100,000$
for studies of variation in the fine-structure constant.

In addition to the increased sensitivity to weak lines, gained at
higher resolution, there is reason to expect we should see
substructure on {\kms} scales.  Distinct small-scale structures are
seen in the interstellar medium of the Milky Way on these velocity
scales \citep{points}.  Very narrow components, indicating cold clouds,
are seen through 21-cm absorption in some damped {\Lya} systems \citep{lane00}.  Also, detailed models of
transitions detected in DLA systems indicate the presence of a cold
phase \citep{wolfe03}.  The large
{\Mgn} to {\Mgi} ratios in many components of strong {\MgII}
absorbers, most of which are produced by lines of sight through $L^*$
galaxies, could be explained by a cold phase that produces the bulk of
the {\MgI} absorption \citep{cwcvogtchar03}.
In the case of the $z=0.9902$ system toward PG$1634+706$, \citet{z99} proposed cold clouds with Doppler parameters of $b(Mg)
\sim 0.75$~{\kms} to explain the strong {\MgI} absorption, but these
could not be resolved at FWHM$\sim7$~{\kms}.  Finally, VLA observations
of Milky Way high velocity clouds indicate that they have $T<900$~K,
and most Milky Way lines of sight show narrow {\CaII} lines from HVCs \citep{richter05}.  However, at FWHM$\sim7$~{\kms}
these {\CaII} lines were unresolved. \citet{richter05} suggested that if the Milky Way is typical of spiral galaxy
halos these narrow absorption features should be commonly seen in
the profiles of strong {\MgII} absorbers and Lyman limit systems.

For the brightest quasars it is already feasible to use the
highest resolution settings on 8-meter class telescopes ($\sim$~40 with $m_V < 16$ for non-exorbitant exposure times) to study
quasar absorption line systems.  In this paper, we present an
exploration of the power of $R=120,000$ in studying four {\MgII}
absorbers along the line of sight toward the magnitude $m_V=14.9$
quasar PG$1634+706$.  Our observations were carried out in the service
mode with the HDS on the Subaru Telescope.

The paper begins, in \S~\ref{sec:2}, with a general consideration of
how accurately line parameters can be determined at super-high resolution.
In \S~\ref{sec:3}, we present our Subaru/HDS observations of
several absorption line systems in the spectrum of one of the brightest quasars,
PG~$1634+706$, and discuss the implications of our $R=120,000$ observations
of these systems.  We also show the Milky Way absorption along this
line of sight.  In \S~\ref{sec:4}, we speculate about the progress that
will be made in the study of quasar absorption lines when many quasars
are observed at $R>100,000$.

\section{SIMULATION OF VERY NARROW ABSORPTION LINES}
\label{sec:2}

\begin{figure*}
\figurenum{1}
\epsscale{1.0}
\plotone{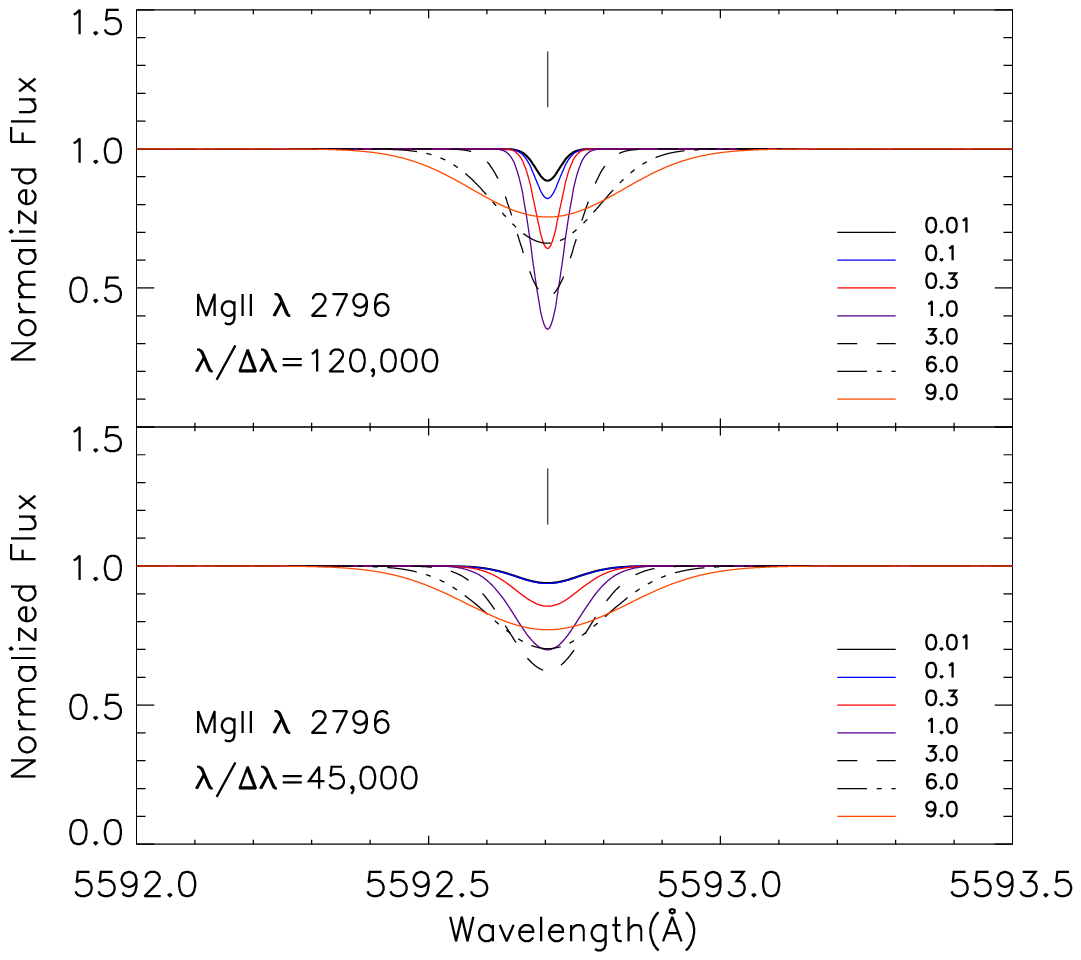}
\protect
\caption{Synthetic absorption profiles of varying Doppler parameter $b$ convolved to spectral resolutions of $R=120,000$ and $R=45,000$. The absorption line is redshifted to $z=1$. The tick marks the center of the absorption profile.}
\label{fig:1}
\end{figure*}

In gas with temperatures T $< 8000$ K, most metals would produce absorption features that are narrower than $b_{therm}$ = 2~{\kms}. Figure~\ref{fig:1} illustrates progressively narrow absorption profiles of {\MgII}$\lambda$ 2796 line with column density $N({\Mgi}) = 10^{12}$ cm$^{-2}$ and with Doppler parameters $b({\Mgi})$ from $9$~{\kms} to $0.01$~{\kms} at high ($R=45,000$) and super-high spectral resolutions ($R=120,000$). The various profiles were produced by convolving a model absorption feature with Gaussian kernels of FWHM = $6.6$ {\kms} and FWHM = $2.5$ {\kms}, to simulate the resolving powers corresponding to the two spectral resolutions. Comparing the profile shapes, it is apparent that, even at high resolution, the width of intrinsically very narrow lines is often misrepresented. A measurement of the $b$ would result in a value significantly higher than the true value, which further affects the estimation of the upper limit for the gas temperature. To explore the effect of resolution further, we evaluate
the consistency with which Voigt profile parameters (i.e., column density and Doppler parameter) can be recovered from very narrow absorption profiles at two resolutions; $R=120,000$ and $R=45,000$ in synthetically simulated spectra. Both these parameters are important for deriving ionization structure, gas phase
metallicities, and other physical conditions for the absorber. In addition, using this approach, we also consider two other effects: (1) blending of lines, and (2) isotope shifts and hyperfine splitting in absorption lines. We evaluate the advantage of increased resolution in recognizing these effects and also in determining how they influence Doppler parameter measurements of absorption profiles. 

\subsection{Very Narrow Single Component Lines}
\label{sec:2.1}

For our simulations we synthesized {\MgII} $\lambda$~2796 lines with two sets of values: (1) column density of $N({\Mgi})$ = $10^{12}$ cm$^{-2}$ and $b({\Mgi})$ = $2.0$~{\kms}, corresponding to a rest-frame equivalent width limit of $W_r(2796)=0.03$~{\AA}, and (2) $N({\Mgi})$ = $10^{12}$ cm$^{-2}$ and $b({\Mgi})$ = $1.0$~{\kms}, corresponding to $W_r(2796)=0.02$~{\AA}. The former characterizes the typically measured profile parameter for the unique class of weak metal line absorbers known as ``single-cloud weak {\MgII}'' systems \citep{weak1, weak2, weak1634}. The latter would be representative of even colder phases of gas ($T<1460$~K) at low ionization.  Our choice of these values for $b({\Mgi})$ for the simulations was also motivated by a desire to find the minimum Doppler width at which we expected to resolve lines at $R=45,000$, and to show the improvement with $R=120,000$.  Thus the simulations are driven by the characteristics of the particular instruments that we compare.

\begin{figure*}
\figurenum{2}
\epsscale{0.8}
\rotatebox{90}{\plotone{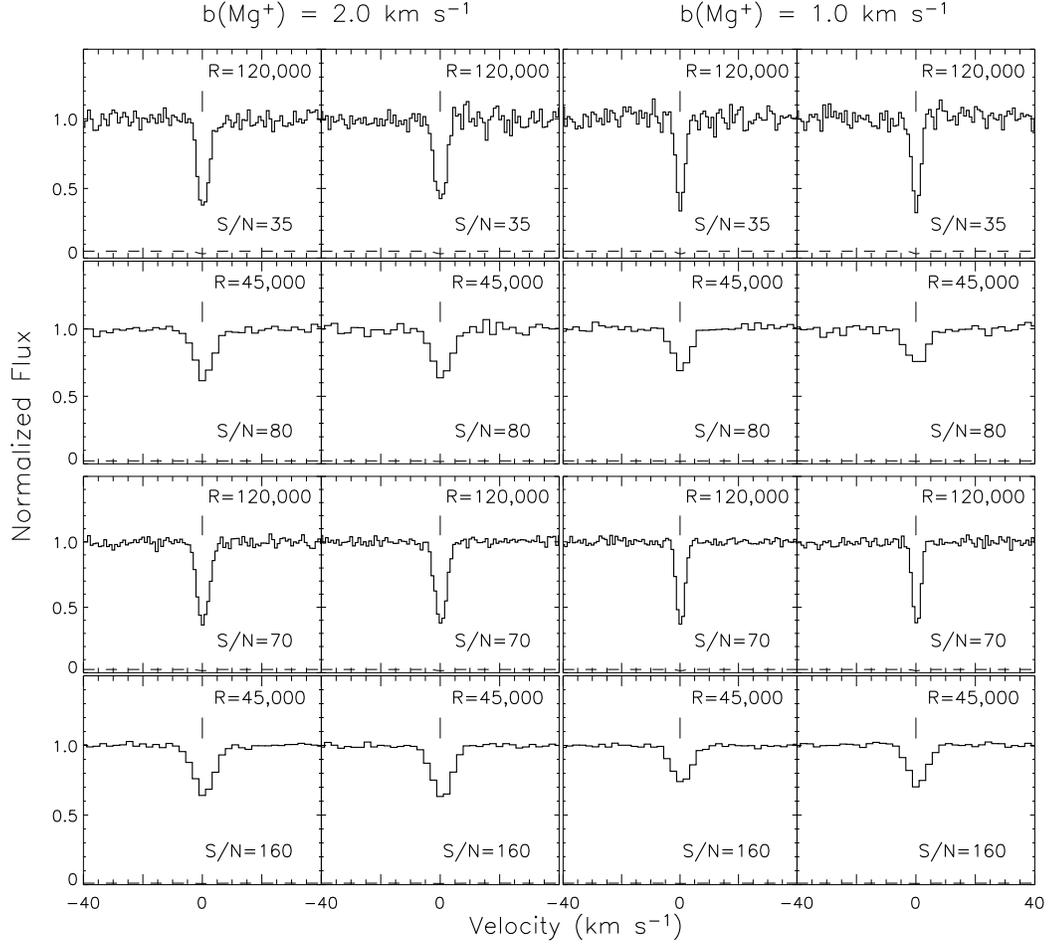}}
\protect
\caption{A sample from our synthetically simulated {\MgII} $\lambda$2796~{\AA} absorption lines with noise added. The two set of lines with Doppler parameters $b({\Mgi})=2.0$~{\kms} and $b({\Mgi})=1.0$~{\kms} have a column density of $N({\Mgi})=10^{12}$~cm$^{-2}$. The dotted line in each panel is the error spectrum. The tick marks represent the center of each absorption profile.}
\label{fig:2}
\end{figure*}

In generating a synthetic spectrum, an ideal absorption line for the chosen parameters was first simulated using a
Voigt function to represent the dependence of absorption on
wavelength. The synthetic profile was initially oversampled to assure accurate representation of narrow lines.  It was then convolved with a Gaussian instrument kernel, as described above. The convolved profile was then resampled at a rate of three pixels per resolution element to enable comparison with observed spectra for a particular $R$. Finally, Poisson noise was added to the spectra, corresponding to a specified $S/N$ per pixel.  To determine the parameters $N$ and $b$ we performed Voigt Profile fits on the absorption lines in $10,000$ simulated spectra.  Figure~\ref{fig:2} shows a sample from those simulated lines. The fit parameters were derived by first generating an initial model profile using an automated fitting routine \citep[{\it AUTOVP}, ][]{dave96}, and then adjusting parameters to minimize the chi-square of that initial model \citep[{\it MINFIT}, ][]{cwcthesis}, after convolving with the instrumental spread function.

Since we want to evaluate the balance between $S/N$ and resolution,
we consider pairs of equal length exposures with $R=120,000$ and $R=45,000$.
The $S/N$ ratios were calculated using the relation: 
\begin{equation}	
\Bigl(\frac{S}{N}\Bigr)_{45} = \Bigl(\frac{S}{N}\Bigr)_{120} \times \sqrt{\frac{F_{45} \cdot \delta\lambda_{45} \cdot f_{45}(s)}{F_{120} \cdot \delta\lambda_{120} \cdot f _{120}(s)}}
\end{equation}
where $F$ is the expected number of photons incident on the spectrograph
slit, $\delta\lambda$ is the grating dispersion (\AA/pixel), and
$f(s)$\ is the slit throughput for the assumed seeing ($s$). For
simplicity we assume that the incident numbers of photons are the same (i.e,
$F_{45} = F_{120}$), though for real systems this is a function of the
telescope throughput and collecting area. (We note also that this
scaling assumes object-limited observations with negligible
contributions from the background and detector read-noise.) We assume
that the grating dispersions are 0.04\AA\ and 0.016\AA\ for the
R=45,000, and 120,000 spectra, respectively. To calculate the slit
throughput, we consider a seeing of 0.6\arcsec~ (FWHM, with a Gaussian
PSF) with slit widths of 0.861\arcsec~ and 0.3\arcsec~ for the R=45,000,
and 120,000 cases. These slit widths were chosen because they represent the two instrumental set-ups used in our observations of PG~$1634+706$ with Keck I/HIRES and Subaru/HDS. In our $R=120,000$ observation of PG~$1634+706$ using Subaru/HDS we were able to obtain a $S/N \sim 55$ pixel$^{-1}$ for a 1 hour exposure under 0.6\arcsec~ seeing condition. Using the relation described above, this scales to a $S/N$ of 128 pixel$^{-1}$ for a spectral resolution of $R=45,000$ for identical exposure time and seeing condition. 

\begin{figure*}
\figurenum{3}
\epsscale{0.8}
\rotatebox{90}{\plotone{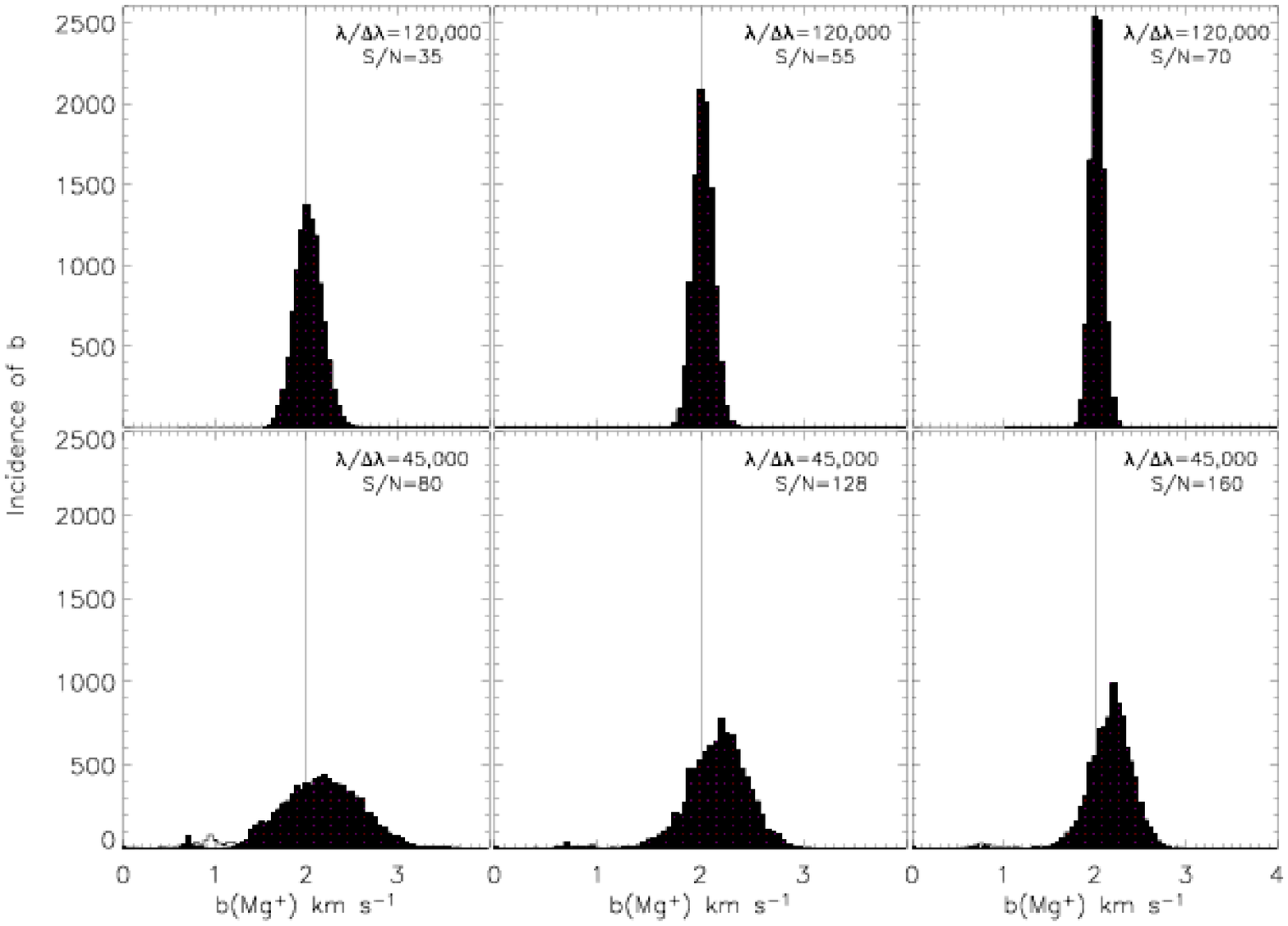}}
\protect
\caption{The distribution of measured Doppler parameter $b$ from $10,000$ realizations of an {\MgII} $\lambda 2796$~{\AA} line with $W_r(2796)=0.03$~{\AA}. The solid historgrams are measurements from well constrained Voigt Profile models that had $\sigma (b) < b$. For each case, the measurements that were affected by pixelization are separately shown as the unfilled histogram. The true value of $b$-parameter is $2.0$~{\kms}}
\label{fig:3}
\end{figure*}

Figure~\ref{fig:3} presents the distributions of the measured values
of $b$ and $N$ for $10,000$ realizations of simulated spectra with $N
= 10^{12}$ cm$^{-2}$ and $b = 2.0$~{\kms} lines for various
combinations of $R$ and $S/N$. Doppler parameter characterizes the
width of the Voigt profile which in turn influences the gas
temperature estimation. Therefore we compare the consistency in the
measured $b$-values for each of the $10,000$ simulated spectra.  For
the $R=45,000$\ simulations, we find the measured $b$-values are
distributed with a large tail to higher values. The mode of this
distribution also falls at a value that is higher than the true
$b$-parameter and it shifts to even higher values as the $S/N$\ ratio
is decreased. In a small subset of cases the Voigt profile models from our
fitting procedure yielded results that have large formal uncertainties,
$\sigma$~$(b)$, in the measurement of $b$. We
find that measurements with $\sigma (b) > b$ are poorly constrained,
often with the data in one or more pixels deviating by more than
3~$\sigma$ from the model fit. These models are shown in the
distribution plots of Figure~\ref{fig:3} as a separate category.  For the $S/N=80$ case, the small peak at $\sim 0.95$~{\kms} results from a pixelization effect.
Typical line profiles with the appropriate sampling are detected in a small discrete number of pixels ($\sim 1$-$2$), which can lead to bias toward particular values of $b$.  For these profiles the noise also sometimes conspires to produce a $b$ value even narrower than the instrumental spread function.  In such a case, the true $b$ value is likely to be
small, but the measurement cannot be claimed as accurate. These simulations therefore reveal the inadequacy of $R=45,000$ data in allowing the recovery of a narrow $b$-value.

In comparision, for the super-high resolution synthetic spectra, the
distribution of the measured $b$ is significantly narrower and the
mode is very close to the true value. Additionally, the intrinsic
value remains the most frequently recovered $b$-parameter even as the
S/N ratio is reduced. For the $S/N$ that are compared, in almost all $10,000$
cases our fitting procedure gave a well constrained measurement of the
$b$-value.  Table~\ref{tab:tab1} lists, for the various simulations
with the two resolutions, the median value of $b$, with asymmetric
$1$-$\sigma$ error bars, and the fraction of Voigt Profile fit models
with accurate fits [$b > \sigma(b)$]. We conclude that at
$R=120,000$, a line profile with a true $b$-value of $2$~{\kms} is
very accurately recovered.  At $R=45,000$, there is a significantly
wider distribution of measurements, with a non-negligible chance of obtaining $b$ significantly greater than $2$~{\kms}, and even occasional measurement of $b\sim1$~{\kms} due to pixelization effects.

\begin{figure*}
\figurenum{4}
\epsscale{0.8}
\rotatebox{90}{\plotone{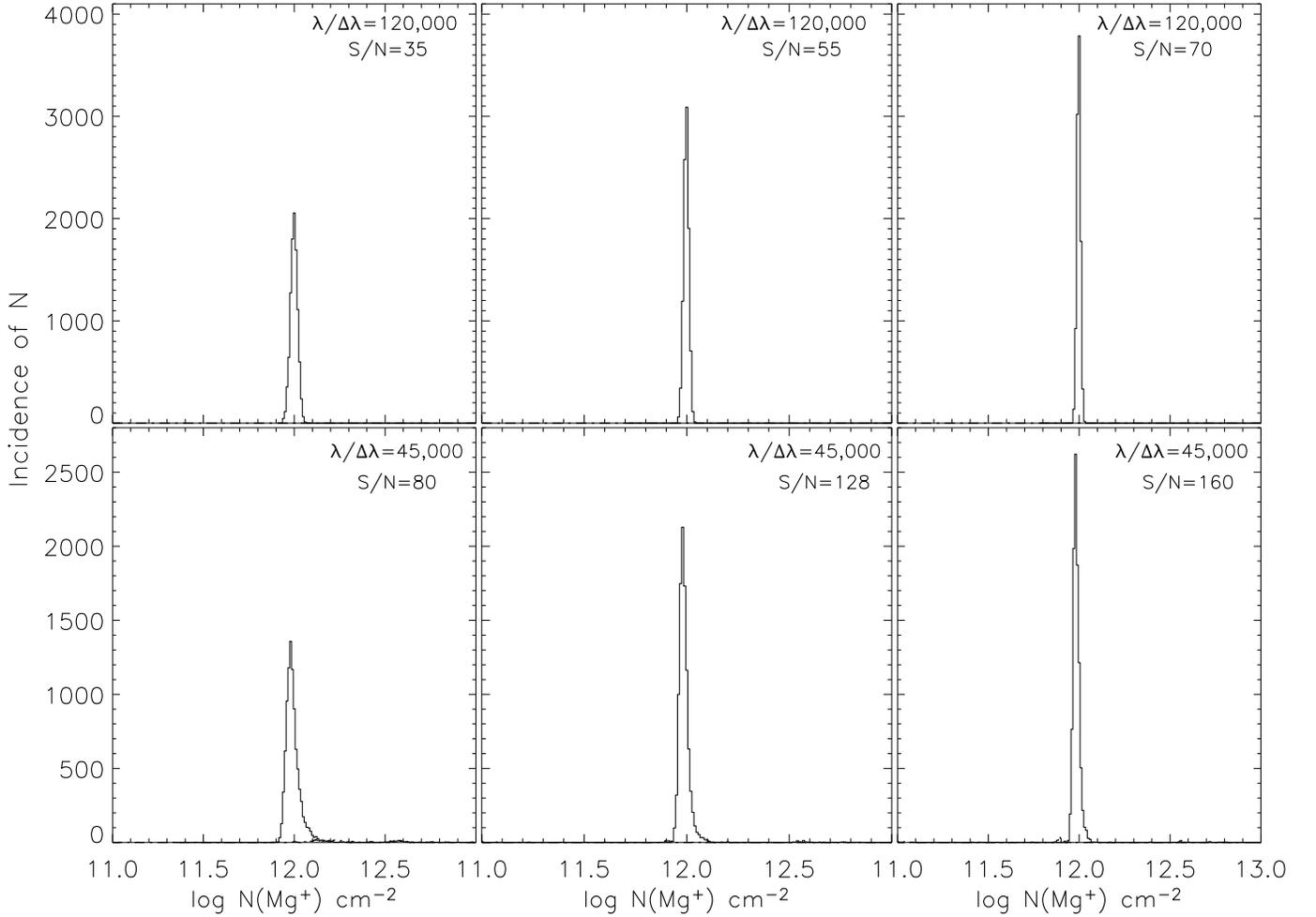}}
\protect
\caption{The distribution of measured $N({\Mgi})$ column density from $10,000$ realizations of the {\MgII}  $\lambda 2796$~{\AA} line with $W_r(2796)=0.03$~{\AA}. The true value of column density is $N({\Mgi})=10^{12}$ cm$^{-2}$.}
\label{fig:4}
\end{figure*}

Figure~\ref{fig:4} shows the distribution of the column density
measurements for the same set of simulations.  The distributions are
broader for the $R=45,000$\ case, with mode values smaller
than the true value.  Measurements thus preserve equivalent width in
the sense that a line with an $N$ measurement smaller than the true
value will have a measured $b$ larger than the true value.  The Voigt
profile models that gave large $\sigma (b)$ also yield large
uncertainties in column density $\sigma$~$(N)$.  Table~\ref{tab:tab1} also summarizes our measurements of $N$-values for the various simulations.  Comparing
values in Table~\ref{tab:tab1}, it is evident that the constraints
from these model fits are poor. We conclude that there is a benefit of
observing at $R=120,000$ for lines with $b=2$~{\kms}, however it is
likely that even at $R=45,000$, the derived value will be roughly
correct.

\begin{figure*}
\figurenum{5}
\epsscale{0.8}
\rotatebox{90}{\plotone{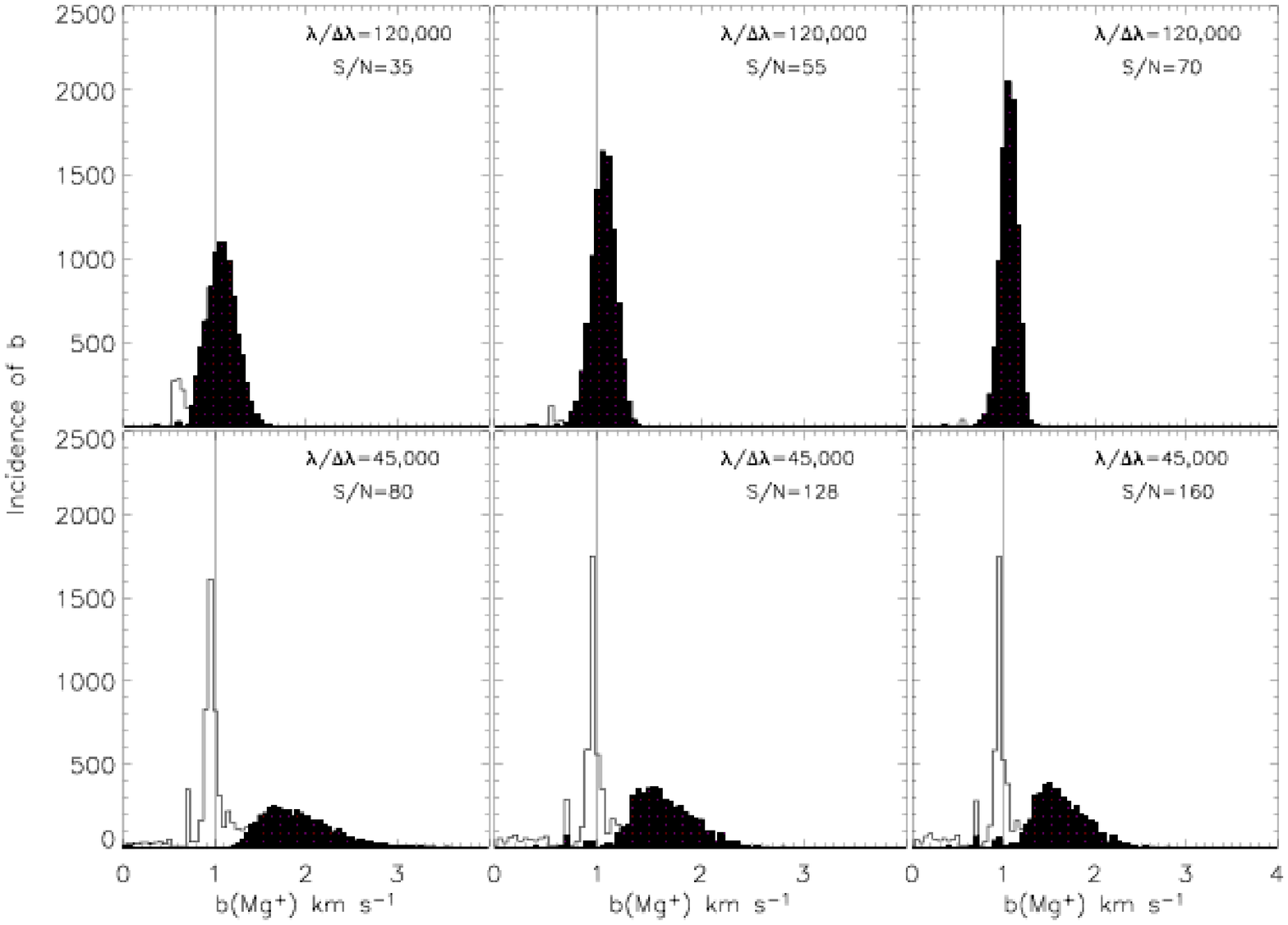}}
\protect
\caption{The distribution of measured Doppler parameter $b$ from $10,000$ realizations of an {\MgII} $\lambda 2796$~{\AA} line with $W_r(2796)=0.02$~{\AA}. The solid historgrams are measurements from well constrained Voigt Profile models that had $\sigma (b) < b$. For each case, the measurements that were affected by pixelization are separately shown as the unfilled histogram. The true value of $b$-parameter is $1.0$~{\kms}.}
\label{fig:5}
\end{figure*}

The real advantage of higher resolution is for even narrower lines.
We performed a similar series of simulations for {\MgII}~$\lambda$2796
lines with $N = 10^{12}$ cm$^{-2}$ and $b=1$~{\kms}.  The
results for the distributions of $b$ parameters for various $S/N$
values at $R=120,000$ and $R=45,000$ are presented in
Figure~\ref{fig:5}. The large spread in the measured value of $b$ for
$R=45,000$ is indicative that this spectral resolution is inadequate
to faithfully recover line widths of narrow line profiles.  We again,
for $R=45,000$ find a large number of measurements with large
$\sigma(b)$ values, for measured $b$ of $\sim 0.95$~{\kms}.  This is
again explained by a pixelization effect, but in this case the
measurement happens to be very close to the true value.  Even if we
had chosen a true value of $b$ between $1$ and $2$~{\kms}, this peak
would still be at $b \sim 0.95$~{\kms}.  Nonetheless, a measurement of
a small $b$ value does imply a true narrow line--width, even if the
measured value is biased by the pixelization.

\begin{figure*}
\figurenum{6}
\epsscale{0.8}
\rotatebox{90}{\plotone{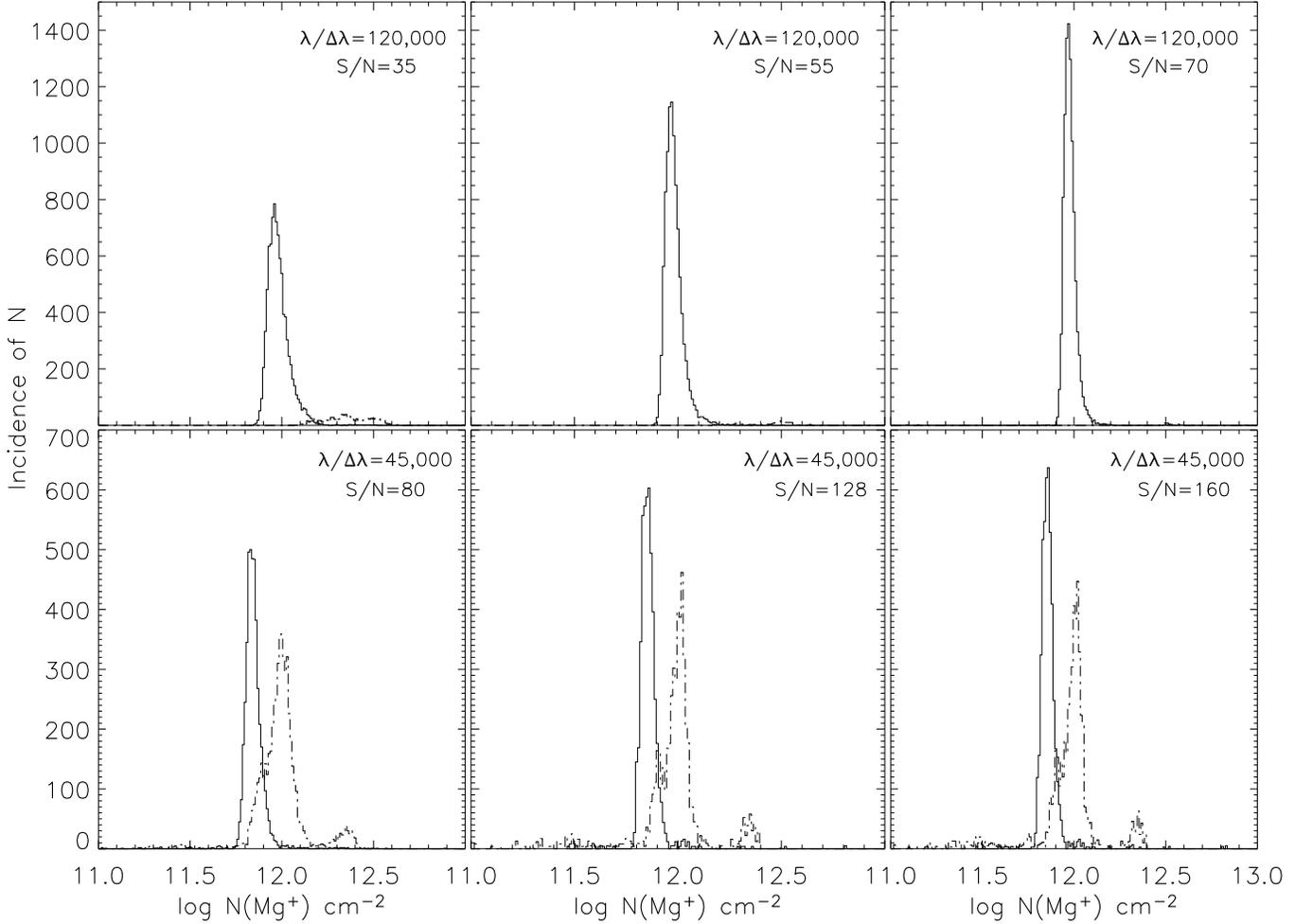}}
\protect
\caption{The distribution of measured $N({\Mgi})$ column density from $10,000$ realizations of the {\MgII}  $\lambda 2796$~{\AA} line with $W_r(2796)=0.02$~{\AA}. The histogram shown in dotted line are measurements that yielded poor constraints ($\sigma (b) > b$). The true value of column density is $N({\Mgi})=10^{12}$ cm$^{-2}$.}
\label{fig:6}
\end{figure*}

Even for the $R = 120,000$ spectra, the mode of the distribution for
measured-$b$ is at a value slightly larger than the true $b$
although in general the measurements are well constrained.
Table~\ref{tab:tab1} lists the median values of Doppler parameters and
column densities, and their $1$-$\sigma$ errors, for various $S/N$
values at the two resolutions. The distributions of $N$-values are
compared in Figure~\ref{fig:6}.

The significance of increased resolution is clearly seen from
Figures~\ref{fig:3} through \ref{fig:6}. The determination of $b$ is also sensitive to the sampling, where $b$ greater than roughly a pixel is required.  In comparison, the precision in measurements from lower resolutions can be improved only
by using substantially higher $S/N$ spectra thus requiring very long exposure time. Furthermore, we find that similar equivalent
width limits are attained in the $R=120,000$ and $R=45,000$ spectrum for equal exposure lengths though the $S/N$ ratio of the former is substantially lower. For example, the $5\sigma$ equivalent width limit for $R=45,000$, $S/N=80$ and $R=120,000$, $S/N=35$ is $5$~m\AA/pixel. Also, as we discuss in the following sub-section and in \S~\ref{sec:3}, observations with $R=120,000$, corresponding to a resolution of $2.5$~{\kms}, are very important for resolving lines and separating velocity sub-structures.

\subsection{Blending of Lines}
\label{sec:2.2}

\begin{figure*}
\figurenum{7}
\epsscale{0.8}
\plotone{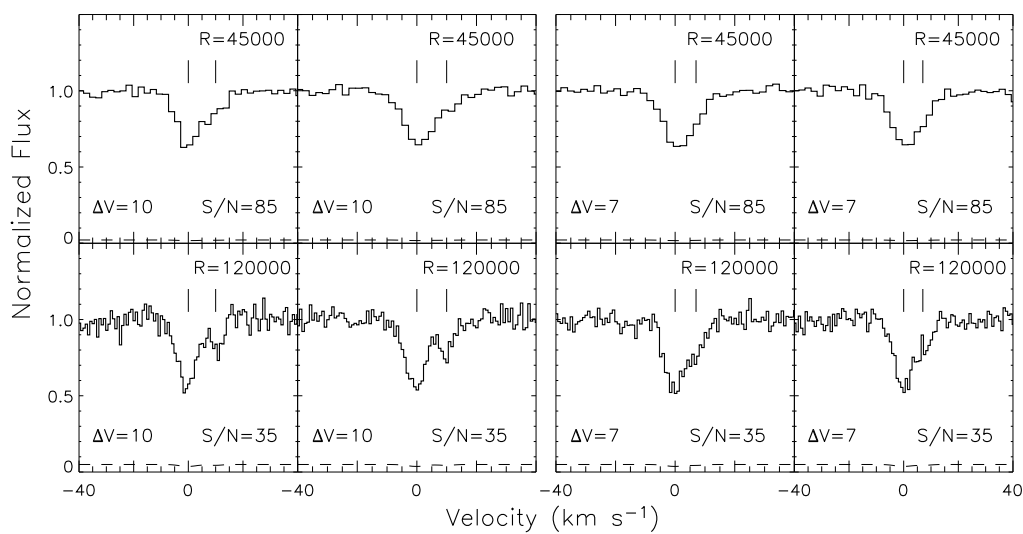}
\protect
\caption{A sample showing closely blended synthetically simulated {\MgII}~2796~{\AA} lines with noise added corresponding to a $S/N$ ratio. The sub-panels in the left panel correspond to the two absorbing components separated in velocity by $10$~{\kms}. The right panel are sample simulated spectra illustrating the case where the two absorbing components are separated in velocity by $7$~{\kms}. The ticks mark the position of the two components. The $S/N$ ratio for the two resolutions are estimated for equal exposure times.}
\label{fig:7}
\end{figure*}

To probe the nature of small scale structure in absorbers it is essential to resolve closely separated velocity components. Here we focus on making estimates of the scale at which superhigh resolution offers a real improvement over intermediate resolution spectra. We chose a simple scenario of two absorbing clouds, with different physical properties, marginally separated in velocity. The column densities and Doppler parameters of the simulated lines were selected to match closely with those of the weak multiple cloud system at $z=1.0414$ in the spectrum of PG~$1634+706$, so that our choice is not completely arbitrary.  This multiple cloud system is discussed in \S~\ref{sec:3.3}. The two components have column densities $N({\Mgi})$ of $10^{12}$ and $10^{11.5}$~cm$^{-2}$ and Doppler parameters $b({\Mgi})$ of 4.0 and 3.0~{\kms}, respectively. This corresponds to an equivalent width ratio of approximately $3:1$ between the two components. In a set of $1000$ simulations with a $S/N$ of $80$ pixel$^{-1}$ at $R=45,000$, and with the two clouds separated in velocity by $\Delta$v=10~{\kms}, we found that in $\sim 84\%$ of the cases Voigt Profile fitting was able to distinguish the two components in the absorption profile. For the same $S/N$ ratio, with the velocity separation reduced to $\Delta$v=7{\kms}, in $\sim95\%$ of the cases the simulated line was fit with just a single broad Voigt profile with a mean value of $\bigl<b$(Mg$^+$)$\bigl>=6.3$~{\kms}. This translates to a larger upper limit for the estimated temperature of the gas (T $< 58000$~K) compared to the true value of T $< 20000$~K. For lines with the same intrinsic properties, but simulated at $R=120,000$ and $S/N=35$ pixel$^{-1}$ (corresponding to an identical exposure time), we found that Voigt profile fits recovered both kinematic components in $\sim97\%$ of the cases with a velocity separation of $\Delta$v=7~{\kms}, and in $100\%$ of the cases with the velocity separation increased to $\Delta$v=10~{\kms}. Furthermore, increasing the $S/N$ ratio of the intermediate resolution spectra did not compensate for the lower resolution.  For example, when the two components were separated in velocity by $7$~{\kms}, even with $S/N=160$ pixel$^{-1}$ at $R=45,000$, the kinematically separate components were resolved only in $\sim40\%$ of the cases. Figure~\ref{fig:7} shows a sample of the simulated lines at the two resolutions. From these results we infer that even at fairly high $S/N$ ratios, intermediate resolutions can be completely inadequate in revealing closely blended velocity structures on scales smaller than $8$~{\kms}. Further, such unresolved lines can lead to overestimation of the temperature of the absorbing gas. More generally, we conclude that, if two lines are narrower than their velocity separation, they can be resolved by a Voigt profile fit if the resolution is less than $\sim70$\% of the velocity separation.  The exact criterion for separating blends is dependant on the quality of the spectra and the Doppler parameters of the two lines.

\subsection{Contribution from Isotopes and Hyperfine Splitting}
\label{sec:2.3}

\begin{figure*}

\figurenum{8}
\epsscale{0.8}
\rotatebox{90}{\plotone{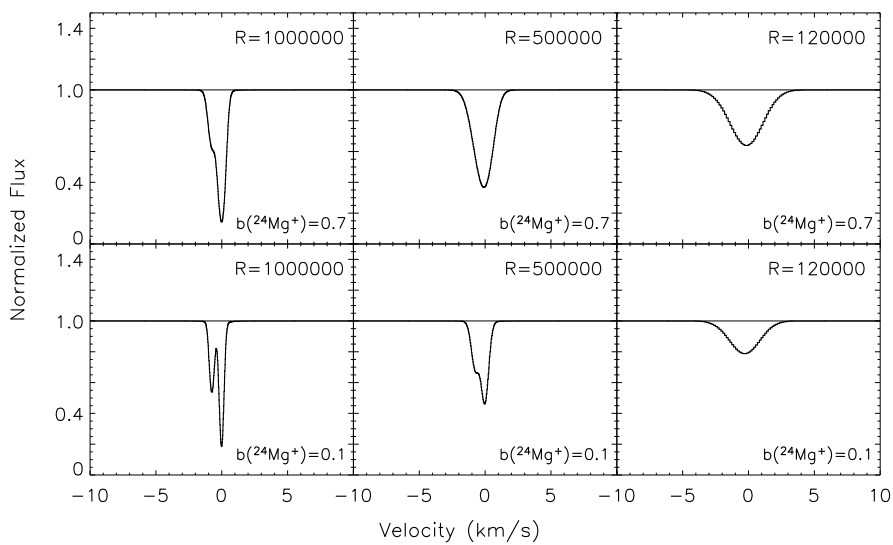}}
\protect
\caption{Synthetic spectra involving isotope and hyperfine lines, without adding Poisson noise. The {\MgII} line is redshifted to $z=1$. A column density of $N({\Mgi})=10^{11.5}$~cm$^{-2}$ was assumed. The $b$ parameter is quoted in units of {\kms}. The line asymmetry because of isotopic line shift becomes distinguishable at spectral resolution $R>500000$ and also when the Doppler width is smaller.}
\label{fig:8}
\end{figure*}

\begin{figure*}
\figurenum{9}
\epsscale{0.45}
\rotatebox{90}{\plotone{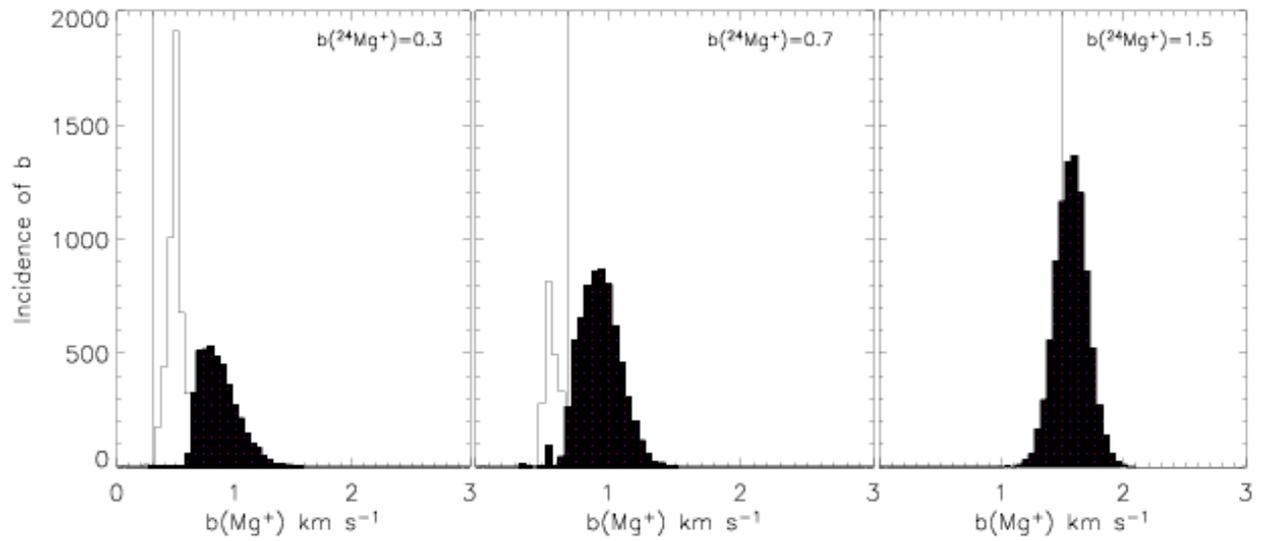}}
\protect
\caption{The distribution of measured Doppler parameter from 10,000 realizations of {\MgII}~2796~{\AA} involving the isotopic and hyperfine lines of Magnesium. The unfilled histograms are measurements with poor constraints ($\sigma(b)>b$). The vertical line shows the true $b$-value assigned to 
${\Mg24i}$ in each of the three cases. The $b$ parameter is quoted in units of {\kms}.}
\label{fig:9}
\end{figure*}

Magnesium has three isotopes with mass numbers 24, 25 and 26 a.m.u, with relative abundances of $78.99\%$, $10.00\%$, $11.01\%$, respectively. The wavelengths of the resonance lines for these isotopes differ from one another. Furthermore, $^{25}$Mg shows hyperfine splitting. The relative shifts in rest frame wavelength for each line of the resonance doublet are measured to be $\lambda_{24} - \lambda^{+}_{25} = 0.0014$~{\AA}, $\lambda_{24} - \lambda^{-}_{25} = 0.0061$~{\AA}, and $\lambda_{24} - \lambda_{26} = 0.008$~{\AA} \citep{morton03}. In velocity space this corresponds to $0.1500$~{\kms}, $0.6540$~{\kms} and $0.8577$~{\kms}, respectively. For such narrow velocity separations, the effect of isotope shifts and hyperfine splitting becomes distinguishable only at extremely high spectral resolutions ($R > 500,000$), as illustrated in Figure~\ref{fig:8}. To estimate the scale at which this effect is likely to influence our $b$-value measurements, we simulated very narrow synthetic lines with true $b({\Mg24i})$ of 0.3, 0.7, and 1.5~{\kms}.  The $b$-values of the other isotope lines were thermally scaled. The simulated lines had $S/N=35$ pixel$^{-1}$ at $R=120,000$. Figure~\ref{fig:9} shows the dispersion in the incidence of measured $b$ for the various cases. Of the $10,000$ simulated lines with $b({\Mg24i})=0.3$~{\kms}, only $\sim 10\%$ of the measurements yielded a $b$-value less than $0.5$~{\kms}, with the recovered $b$-value almost always larger than the true value. For a higher true $b$-value of $0.7$~{\kms}, the measurements were still distributed across a range from $0.7$ to $1.5$~{\kms}. The peak of this distribution, however, was closer to the true-value as compared to the previous case. For the final case, with $b=1.5$~{\kms}, the spectral resolution became adequate to repeatedly measure values close to the true $b$ value along with consistently yielding measurements that are reliable ($\sigma(b)<b$). In this case, although the isotope and hyperfine lines were still contributing to the final absorption profile, their contribution did not induce a significant change in the line width as the individual line widths were now larger than the largest separation between the isotope lines.  

With these results we infer that isotopic line shifts and hyperfine splitting are to be considered as influencing our measurements only when the true line width is really narrow (T~$ < 700$~K, $b({\Mgi})<0.7$~{\kms}). Furthermore, for such narrow {\MgII} lines, identifying the effect would require spectral resolutions of $R>500,000$.

\section{SUPER HIGH RESOLUTION OBSERVATION OF PG~$\mathbf{1634+706}$}
\label{sec:3}

We obtained a super high resolution spectrum of PG~$1634+706$ with the
High Dispersion Spectrograph \citep[HDS]{hds} on Subaru. Using a
0.3\arcsec~slit width, we were able to attain $R=120,000$ spectrum
with a sampling rate of $3$ pixels per resolution element.  Our 1~hr
observation yielded a spectrum with an average $S/N \sim 30$
pixel$^{-1}$ over the entire spectral range. The reduced $1$D spectrum
was extracted in a standard manner with the IRAF software {\footnote
{IRAF is distributed by the National Optical Astronomy Observatories,
which are operated by AURA, Inc., under cooperative agreement with
NSF}}. Wavelength calibration was done using a standard Th--Ar
spectrum. The spectrum was continuum fitted with a third--order cubic
spline function and normalized. Previously identified weak and strong
{\MgII} absorption systems in a $R=45,000$ Keck/HIRES spectrum of
PG~$1634+706$ made this observation an ideal test case. The HIRES data
aquisition and reduction procedure is described in
\citep{cwcvogt01}. The short exposure on HDS reached an equivalent
width limit several times smaller than our exposure of twice its
length with the HIRES instrument at a lower resolution of
$R=45,000$. Details of the two observations are listed in
Table~\ref{tab:tab2}. Figures~\ref{fig:10a}--\ref{fig:10c} show the 
wavelength regions that are covered in both the HDS and HIRES
spectrum.

In the following subsections, we compare detections of the strong,
single cloud weak , and multiple cloud weak {\MgII} absorption
features in the super-high resolution HDS and the high resolution
HIRES spectra for this quasar.  We also consider the Milky Way
absorption along this line of sight.  We discuss the particular case
of these systems, but also generalize to discuss how studies of the
same type of system would be affected by higher resolution
observations. Since the scope of this paper is to compare observations
taken at different spectral resolutions, we defer detailed analyses of
the absorption-line systems presented herein to future work.

\subsection{Strong {\MgII} absorber at $\mathbf{z=0.9902}$}
\label{sec:3.1}

\begin{figure*}

\figurenum{11}
\epsscale{0.9}
\plotone{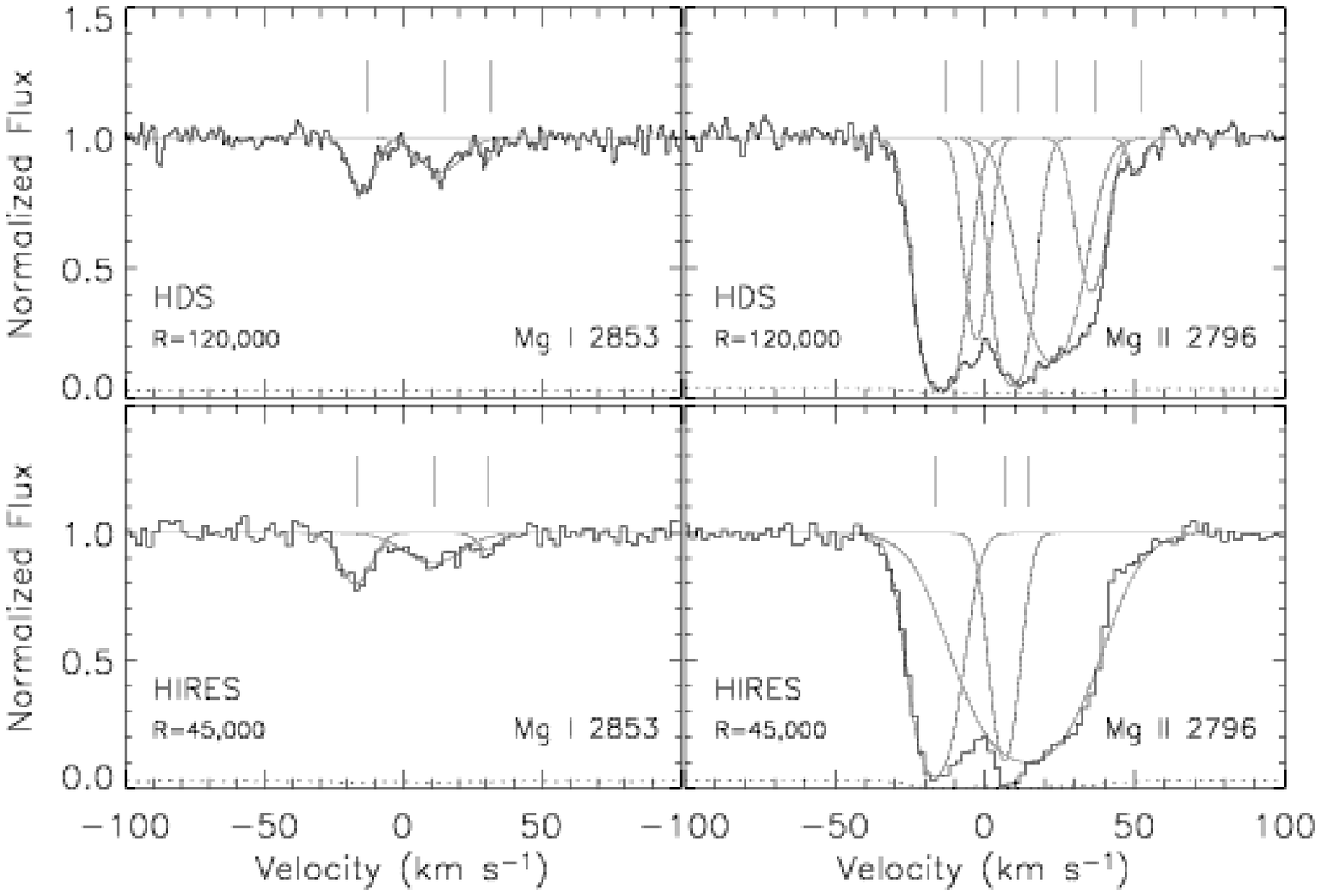}
\protect
\caption{The $z = 0.9902$ strong {\MgII} system as captured by the Keck/HIRES and Subaru/HDS spectrum of PG~$1634+706$. The left panel shows the low ionization {\MgI} feature associated with this strong system. The thin solid line superimposed on the spectrum are the various Voigt Profile components contributing to the absorption feature. The tick marks placed above the features represent the center of each component. The dotted line is the error spectrum. The results from the Voigt Profile fit are listed in Table~\ref{tab:tab3}}
\label{fig:11}
\end{figure*}

The $z=0.9902$ system along the line of sight to the quasar
PG~$1634+706$ has previously been studied using a combination of
the $R=45,000$ Keck/HIRES spectrum, data from HST/FOS
\citep{archive2,jane00} and from HST/STIS \citep{z99}.
\citet{cwcvogtchar03} found that this strong {\MgII} doublet could be
adequately fit with five blended components. With the higher resolving
power of HDS, additional kinematic components were revealed in the
{\MgII} absorption feature. Our Voigt profile fit decomposes the
{\MgIIdblt} profile into seven components. Figure~\ref{fig:11} shows
the HIRES and HDS data, comparing differences in the profile shapes
for the two resolutions. The asymmetrical shapes for the
{\MgII}$\lambda$2796 line at $\Delta v \sim 22$~{\kms} and $\Delta v
\sim 8$~{\kms} from the flux--weighted system center in HIRES spectrum
were indicative of narrowly blended features which are clearly
distinguished in the profiles in the HDS spectrum.
Table~\ref{tab:tab3} presents our Voigt fitting parameters for
{\MgIIdblt} from the Keck/HIRES and the Subaru/HDS observations.
This ability of superhigh resolution to resolve additional components
in strong metal line absorbers was also noted by \citet{chand06}.

Strong {\MgII} absorption systems usually have absorption from gas in
different phases (different densities and temperatures) residing
within an impact parameter of 35h$^{-1}$ kpc of a luminous galaxy with
$L>0.05$~L$^{*}$ \citep{bb91,steidel97}. Dissecting the {\MgII}
absorption profile to reveal kinematic structures at small velocity
separations of $\Delta v < 1$~{\kms} is therefore useful in developing
a physical understanding of the spatial distribution of gas in
galaxies along the line of sight. If the observed {\MgII} absorption
signatures are characteristic of disk ISM \citep{lanzettabowen,
charcwc98}, then measuring structure on the smallest scales is
important to determine if the interstellar medium of high redshift
galaxies follows a complex fractal structure similar to Milky Way
\citep[e.g.,][]{elmegreen}. Resolving closely blended components, as
in the $z \sim 0.9902$ system is also essential in developing
ionization models that provide strong constraints on physical
parameters.  This is especially relevant when there is a difference or
gradient in metallicity, abundance pattern or ionization from cloud to
cloud.

\citet{cwcvogtchar03} studied $23$ {\MgII} systems at $z\sim1$ and found
that many strong {\MgII} systems have observed column density ratios
of N({\Mgn})/N({\Mgi})~$ > 0.01$. This value is the maximum expected
if the absorption is taking place in gas at a single phase under
photoionization equilibrium. In developing a model for this $z=
0.9902$ absorption system based on the $R=45,000$ Keck/HIRES
observations, \citet{z99} found that the {\MgI} absorption line in this system is too
strong to arise in the same phase as {\MgII} for all possible ionization
parameters. To reconcile this discrepancy, \citet{z99} invoked an
additional lower ionization phase that gave rise to the majority of
{\MgI} absorption, but which produced small amounts of {\MgII}.  The
lower ionization phase also produced most of the neutral hydrogen
absorption from this system, with a range $17.3 < \log N({\Hn}) <
17.7$ for the five {\MgI} clouds. In this model, the {\MgI} phase gas
has a very low temperature ($T \sim 800$~K, $b \sim 0.75$~{\kms}),
high density (n$_H \sim 200$ cm$^{-3}$) and small size ($\sim 100$
AU), reminiscent of a cold pocket embedded in a warm, ionized
inter-cloud medium as in the ISM of galaxies.  The spectral resolution
of HIRES was inadequate to show the narrow line widths that we might
expect from such cold, small-scale clumps of gas.  However, the
present Subaru/HDS spectrum provides a direct test of the \citet{z99}
prediction for this $z=0.9902$ absorber.

\begin{figure*}
\figurenum{12}
\epsscale{0.6}
\rotatebox{90}{\plotone{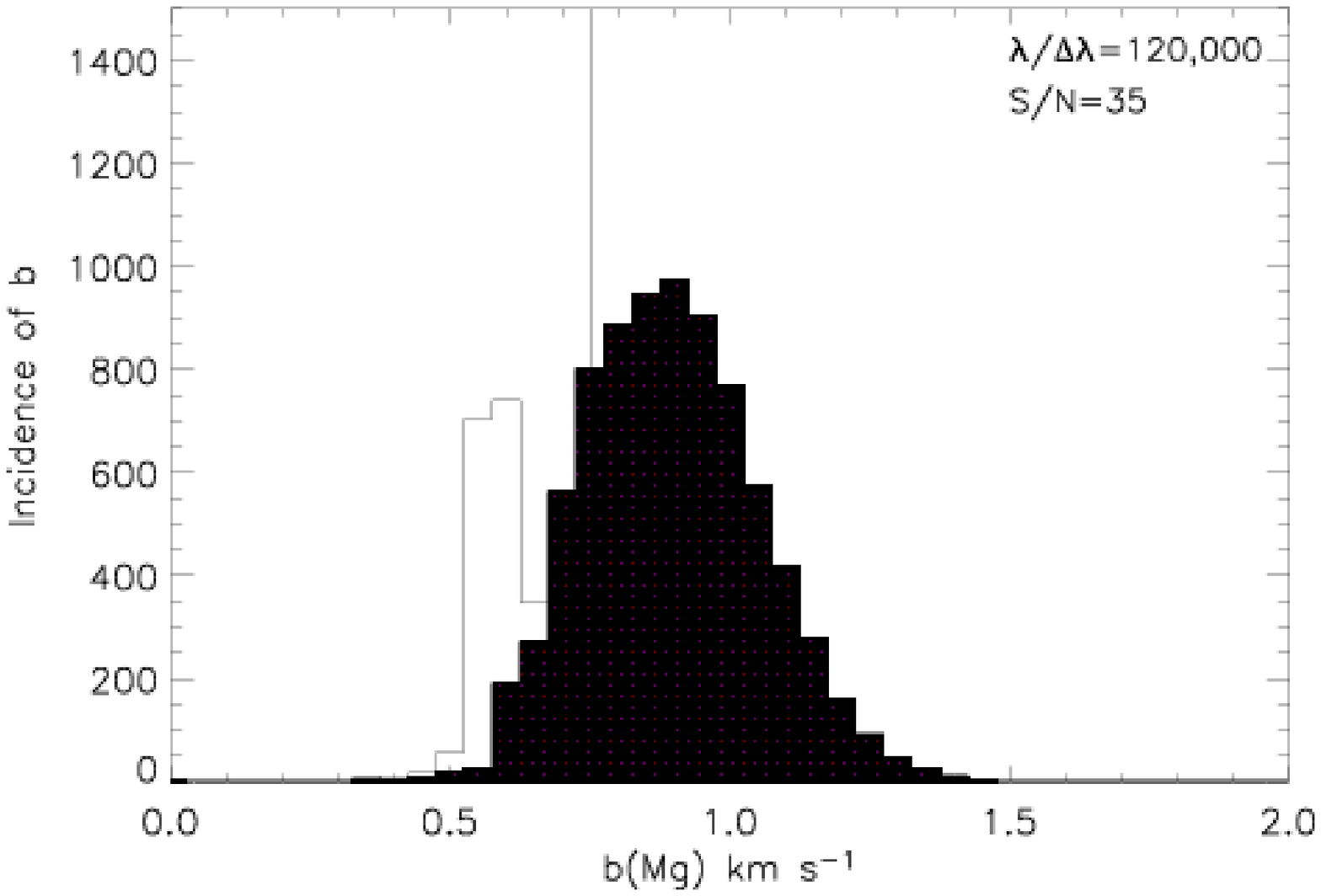}}
\protect
\caption{The distribution of measured $b$ parameter from 10,000 realization of {\MgI}~2853~{\AA} by considering the isotopic lines of Magnesium. The unfilled histogram includes measurements that yielded poor constraints($\sigma(b)>b$). The vertical line shows $b({\Mgl})=0.75$~{\kms}, which is the true Doppler parameter for that line. The line was chosen to have a column density of $N({\Mgn})=10^{11.1}$~cm$^{-2}$ to match with the model predicted by \citep{z99}.}
\label{fig:12}
\end{figure*}

Figure~\ref{fig:12} shows the absorption from {\MgI}~$\lambda$2853 in
the $z=0.9902$ system, as seen in the HDS spectrum.  The Voigt profile
fit constraints on the various line features is listed in
Table~\ref{tab:tab3}. For the three clouds used to fit the {\MgI}
profile, the line widths are not as small as what \citet{z99} had
predicted in their models (see their Table~1). For the line width predicted by their model, the {\MgI} absorbption has to be from gas at temperature $T<800$~K. Our simulation results from \S~\ref{sec:2.3} show that in instances where the true-$b$ value is very small, Voigt profile measurements can yield higher values because of unresolved isotopic lines. To investigate if our measurements of $b$-value for the various components of {\MgI} is influenced by this effect, we resort to simulations. The isotopic lines of {\MgI} occur at $\lambda_{24}=2852.9635$~{\AA}, $\lambda_{25}=2852.9655$~{\AA} and $\lambda_{26}= 2852.9674$~{\AA} \citep{beverini}. The {\MgI}~$2853$~{\AA} line does not have hyperfine structure. In our simulations, we used the profile parameters predicted by the model (N({\Mgn})=$10^{11.1}$~cm$^{-2}$, $b({\Mgn})=0.75$~{\kms}) for the isotopic line of ${\Mgl}$, with the $b$-parameters of the other two isotopic lines thermally scaled. The spectra were simulated for a $S/N=35$ pixel$^{-1}$ at $R=120,000$, with the lines redshifted to match the redshift of the system. Figure~\ref{fig:12} shows the distribution in the incidence of measured $b$ from a simulation of $1000$ lines. It can easily be noticed that, although the measured $b$ is almost always larger than the true $b$ of $0.75$~{\kms}, a measurement of $2$~{\kms} or higher (as in the observed HDS spectrum) never occurs. It is therefore to be concluded that the hypothesis of a two phase structure as an explanation of the {\Mgn} to {\Mgi} ratio
is therefore not consistent with the high resolution observations.  It
is clear from \S~\ref{sec:2} that a component with $b=0.75$~{\kms}
will not be measured to have $b$ as high as several {\kms} at
$R=120,000$.  Therefore, another explanation must be sought for this
system.  \citet{tappeblack} suggested that a UV flux contribution from
the absorbing galaxy and/or Mg $+$ H$^+$ charge transfer could
significantly influence the {\Mgn} to {\Mgi} ratio.  Their
$R=100,000$ VLT/UVES observation of a different system, at $z=0.79089$
toward PKS~$2145+067$, yield $b({\Mgi}) = 1.02\pm1.65$~{\kms} for the
one component detected in {\MgI} in a spectral region with $S/N \sim
30$.  The error is too large to draw a conclusion in that case.
Another possibility for the $z=0.9902$ system is that there are more
components in the {\MgI} than we are able to resolve.

We conclude that our $R=120,000$ spectrum does not provide support for
a cold phase model for the $z=0.9902$ system, though some modified
version could apply.  Direct tests with super high resolution for
other similar systems are needed.  One excellent candidate is the
$z=0.4523$ multiple-cloud, weak {\MgII} system toward the quasar
HE~$0001-2340$\ ($z = 2.28, m_V = 16.7$), which has detected {\FeI}, and
{\CaI} as well as {\MgI} \citep{jones06}.  Cold phase models may also
have significant consequences for the study of damped Lya absorbers
(DLAs) and sub-DLAs. \citet{lane00} have resolved narrow components
($b\sim2-3$~{\kms}) in an {\HI} $21$cm study of the DLA system at
$z=0.0912$ toward B~$0738+313$. {\MgI} lines in these systems are
expected, from thermal scaling, to be significantly narrower.  
Also, more recently, \citet{srianand05} report measuring kinetic temperature
of $\sim 150$~K from H$_2$ lines in three DLAs.  Such a low gas phase
temperature can produce narrow metal lines, with Doppler width of
$\sim 3$~{\kms}, that can be accurately recovered with superhigh
resolution spectra.  A fair
fraction of DLAs have detected molecules \citep[e.g., ][]{petitjean} , indicating directly that they have a cold phase.  Other DLAs and Lyman limit
systems may also be produced by lines of sight through cold regions
surrounding molecular clouds, or regions that did not quite reach the
threshold for star formation.

\subsection{Single-Cloud Weak {\MgII} absorbers}
\label{sec:3.2}

Most single-cloud, weak {\MgII} absorbers (rest--frame
equivalent width $W_r(2796)<0.3$~{\AA}) are not associated with
luminous galaxies, although models designed to fit the Lyman series
line and the observed lack of Lyman break constrain the metallicity of
the absorbing gas to be at least 10$\%$ solar.  Photoionization models
suggest that the structures responsible for weak {\MgII} absorption
might be unstable over astronomical timescales, because of pressure
imbalance between a high density, low ionization ($p/k =
10^{-5}$~cm$^{-3}$~K$^{-1}$) and a low density high ionization ($p/k =
10^{-8}$~cm$^{-3}$~K$^{-1}$) phase
\citep{weak1634,anand05}. The conclusions from photoionization models
are based on the observed values of column density and Doppler
parameter for the low and high ionization transitions. The best
constraints on these parameters were mostly from high resolution data
at $R\sim45,000$ or lower.  The $b$ parameters measured at
$R\sim45,000$ ($b\sim2$ or $3$~{\kms}) allowed for temperatures of
thousands of Kelvin, similar to the temperatures inferred for the high
ionization phases. If the true Doppler parameter is smaller than that derived from the
HIRES measurements, then it would imply that the low ionization phase
has a much lower temperature than the high ionization phase. This
influences the inference about their stability. Line widths narrower
than $\sim2$~{\kms} ($T < 15,000$~K) cannot be accurately measured at
$R=45,000$.  Therefore the temperature estimations for the low
ionization phase in these absorbers may not be accurate.  Our
$R=120,000$ Subaru/HDS spectrum allows us to test the assumption that
these single-cloud weak {\MgII} lines are resolved.

\begin{figure*}

\figurenum{13}
\epsscale{1.0}
\plotone{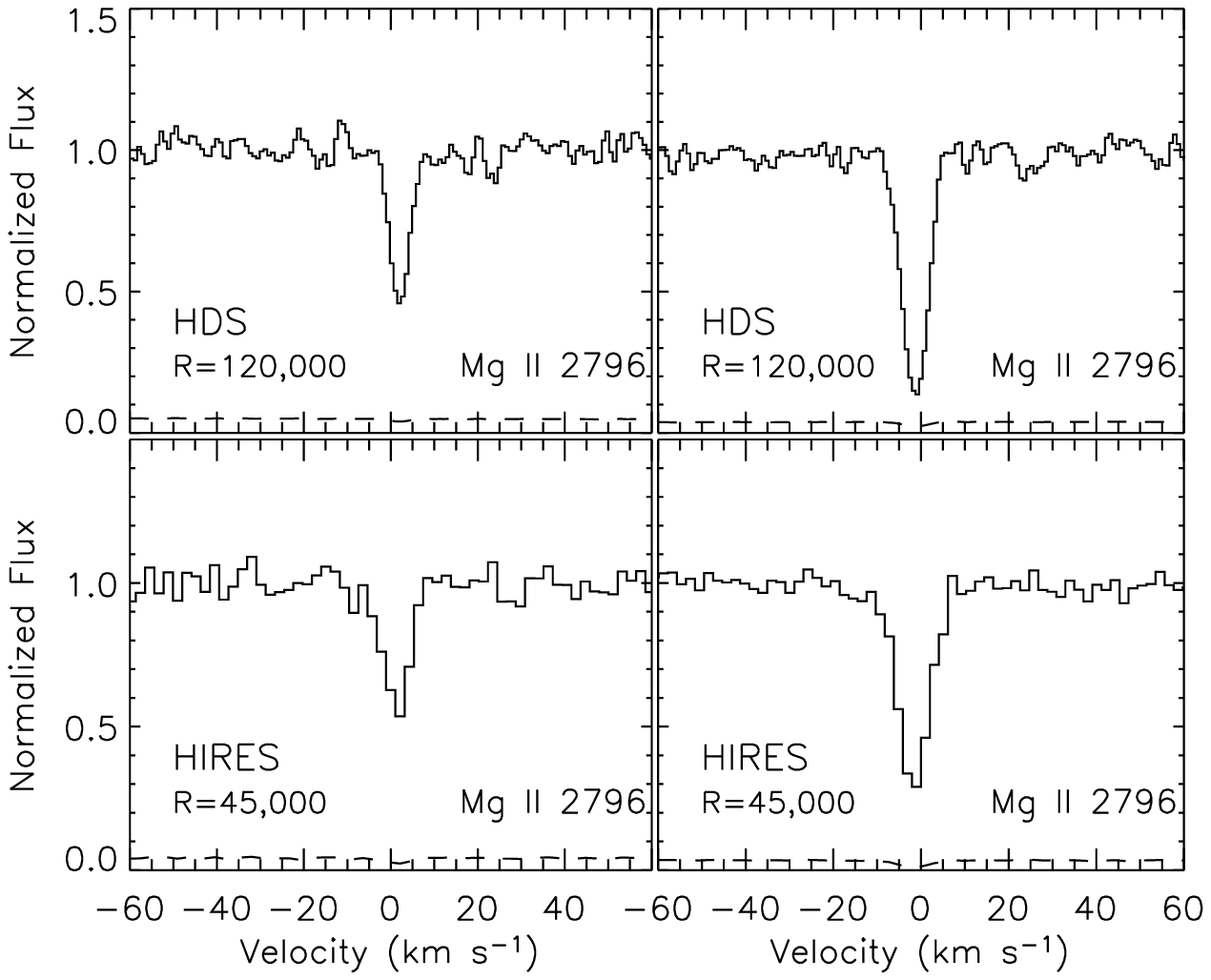}
\protect
\caption{The two single--cloud weak {\MgII} systems at redshifts 0.8181 ({\it left panel}) and 0.9056 ({\it right panel}) as detected in the Keck/HIRES and Subaru/HDS spectrum of PG~$1634+706$. The dotted line is the error spectrum. The Voigt Profile fit results are listed in Table~\ref{tab:tab4}}
\label{fig:13}
\end{figure*}

For the two single cloud weak {\MgII} absorbers at $z=0.9056$ and
$z=0.8181$ found in the PG~$1634+706$ spectrum, we find from the HDS
spectrum that the weak {\MgII} lines are not any narrower than values
measured at R~$45,000$ (Figures~\ref{fig:13}). The column densities and
Doppler parameters from simultaneous Voigt profile fits to {\MgIIdblt}
are listed in Table~\ref{tab:tab4}.  Thus at least for these cases,
the lines were already sufficiently resolved at $R=45,000$ to allow
measurements of the $b$-values that were accurate within the errors.
These measurements are in concordance with the instability inference
based on photoionization model. Analogs of weak {\MgII} absorbers are
known to exist in the present day universe \citep{anand05} as well as
at intermediate \citep[$z\sim1$, ][]{weak1} and higher redshifts
\citep[$z\sim2$, ][]{lynch06}. Therefore confirming the transient
nature of these systems is significant in addressing the question of
the evolution of the processes that created these absorbing
structures.  A larger sample of weak {\MgII} absorbers should be
measured with super high resolution in order to verify that we can
generalize this conclusion.  However, the real value in $R>100,000$
observations in the study of weak {\MgII} absorbers will be the
enhanced sensitivity which will allow us to establish the shape of the
distribution function for the weakest absorbers, and to infer their
physical properties.

\subsection{Multiple-Cloud Weak {\MgII} absorbers}
\label{sec:3.3}

\begin{figure*}
\figurenum{14}
\epsscale{1.0}
\plotone{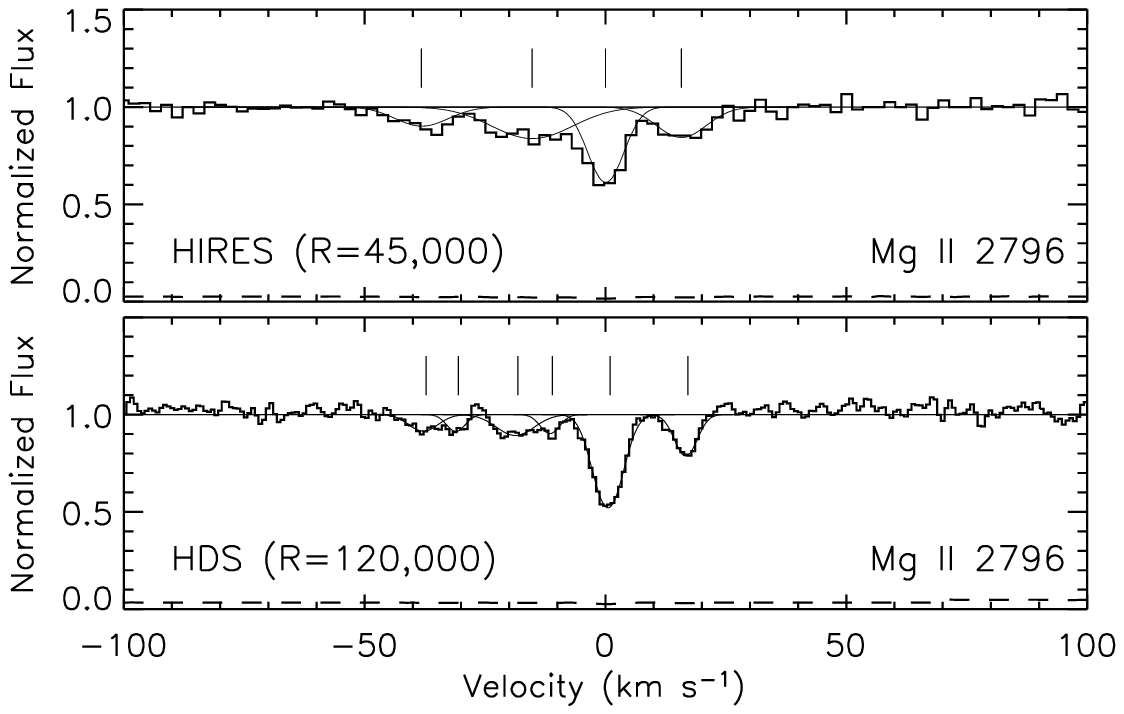}
\protect
\caption{The multiple--cloud weak {\MgII} system at $z = 1.0414$ as detected in the Keck/HIRES and Subaru/HDS spectrum of PG~$1634+706$. The thin solid line superimposed on the spectrum are the various Voigt Profile components contributing to the absorption feature. The tick marks placed above the features represent the center of each component in the absorption profile. The dotted line represents the error spectrum. The Voigt Profile fitting results are listed in Table~\ref{tab:tab4}}
\label{fig:14}
\end{figure*}

\begin{figure*}
\figurenum{15}
\epsscale{1.0}
\plotone{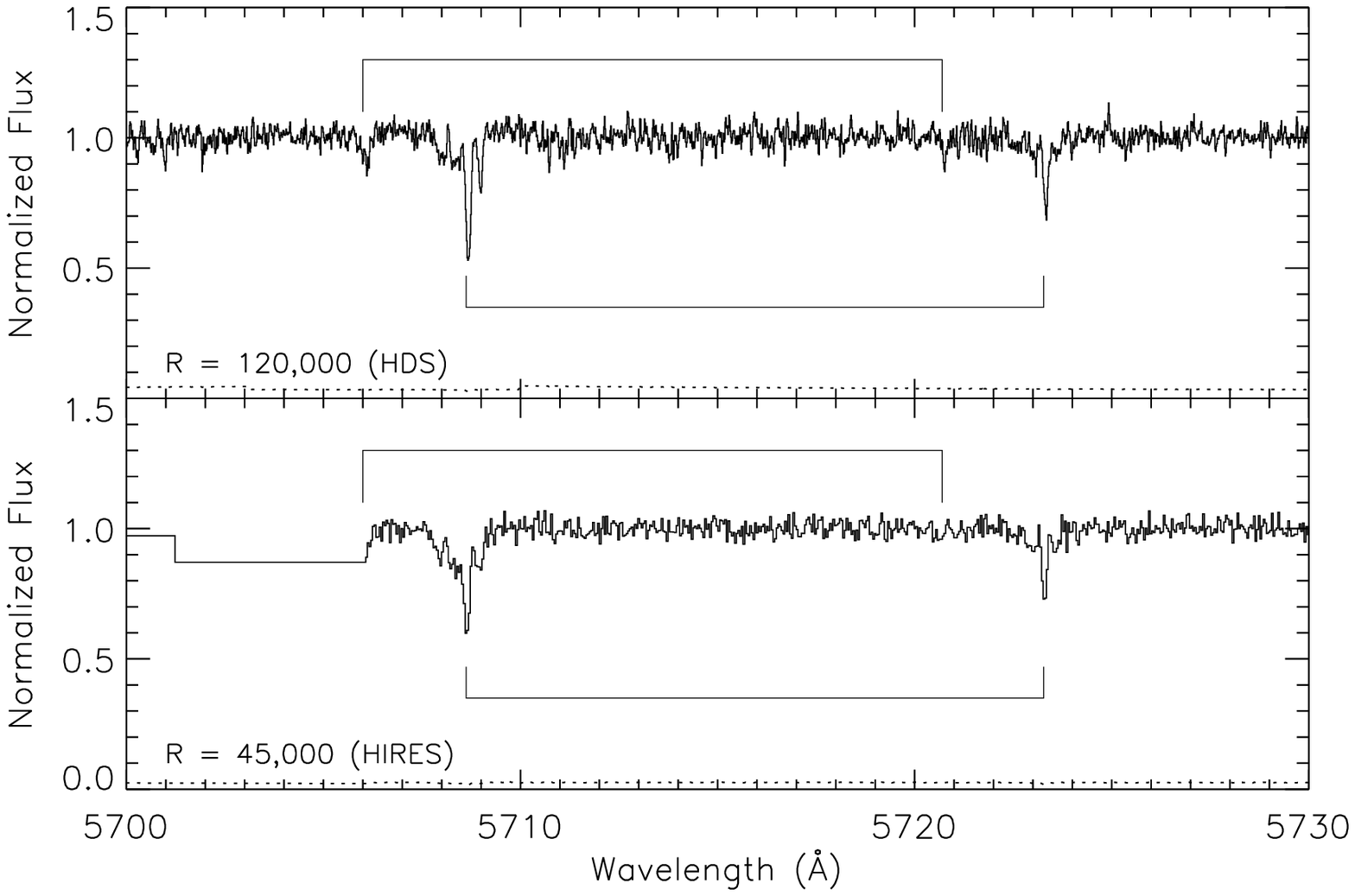}
\protect
\caption{The $z = 1.0414$ multiple--cloud weak {\MgII} system and the associated subsystem in the spectrum of PG~$1634+706$. The dotted line represents the error spectrum. The thin solid lines mark the respective positions for the $\lambda$$\lambda$ 2796 {\AA} and 2803 {\AA} lines that form the doublet pair. The two absorption systems are separated by $\sim 150$~{\kms}. In the HIRES spectrum, for the bluer subsystem, {\MgII} $\lambda 2796$ transition was not covered due to an echelle order break. However, the weaker pair of the doublet, {\MgII} $\lambda 2803$, was not detected down to an equivalent-width limit of $3\sigma$. However, there is a clear detection of this weak subsytem in the HDS spectrum.}
\label{fig:15}
\end{figure*}

Approximately 35$\%$ of weak {\MgII} absorbers have multiple clouds
\citep{weak2}.  Based on their kinematics, these multiple-cloud, weak
{\MgII} absorbers can be described as ``kinematically compact'' or
``kinematically spread''.  The kinematically compact systems have been
hypothesized to arise in dwarf galaxies, and the kinematically spread
systems in the outskirts of spiral galaxies \citep{masiero05,ding05},
but it is likely that the situation is more complicated.  Varying
degrees of line blending in the absorption features implies that the
Voigt profile fit results are sometimes going to be sensitive to
resolution. This is illustrated by the $z = 1.0414$ multiple-cloud,
weak {\MgII} system in the spectrum of PG~$1634+706$.  Based on HIRES
data, this weak {\MgII} absorber was classified as a multiple cloud
weak {\MgII} system with $4$ discrete {\MgII} clouds
\citep{cwcvogtchar03}.  Two additional weak components were resolved in
Voigt profile fits to the higher resolution data in the same region,
as shown in Figure~\ref{fig:14}.  \citet{zonak} produced models for
the system based on the HIRES data along with $R=30,000$ ultraviolet
spectra obtained with HST/STIS.  It is apparent from our HDS spectrum,
in Figure~\ref{fig:15}, that there are two subsystems in {\MgII},
separated by $\sim150$~{\kms}.  For the blueward subsystem
\citep[see][]{zonak}, {\MgII}~$\lambda$2796 was not covered in the
HIRES spectrum, and {\MgII}~$\lambda$2803 was not detected, but
several other metal-line transitions (e.g. {\SiIII}~$1207$, {\SiIVdblt} etc) 
were detected in the HST/STIS
spectrum.  The greater sensitivity of the $R=120,000$ HDS spectrum led
to detection of {\MgII}~$\lambda$2803 in the blueward subsystem, which
was a prediction of the models of \citet{zonak}

Enhanced resolution allows the detailed shapes of absorption profiles
to be surveyed, and blends to be separated.  Some single-cloud weak
{\MgII} absorbers appear asymmetric when observed at $R=45,000$
\citep{weak1}, suggesting they may have more than one Voigt profile
component.  More generally, it is of interest what fraction of the
single-cloud weak {\MgII} absorbers are truly fit by single Voigt
profile components when viewed at higher resolution.  The profiles of
the $z=0.9056$ and $z=0.8181$ systems observed at $R=120,000$ here,
still are fit adequately by a single component, suggesting a very
simple structure.  Observing some asymmetric single-cloud weak {\MgII}
may reveal some to be more similar to kinematically compact multiple
cloud weak {\MgII} absorbers, or they could just be variations of
ordinary single-cloud weak {\MgII} absorbers.  Determinations of their
metallicities and comparisons of kinematics of other transitions
should allow a separation of these classes.

\subsection{Milky Way Absorption and High Velocity Clouds}
\label{sec:3.4}

\begin{figure*}
\figurenum{16}
\epsscale{0.9}
\rotatebox{90}{\plotone{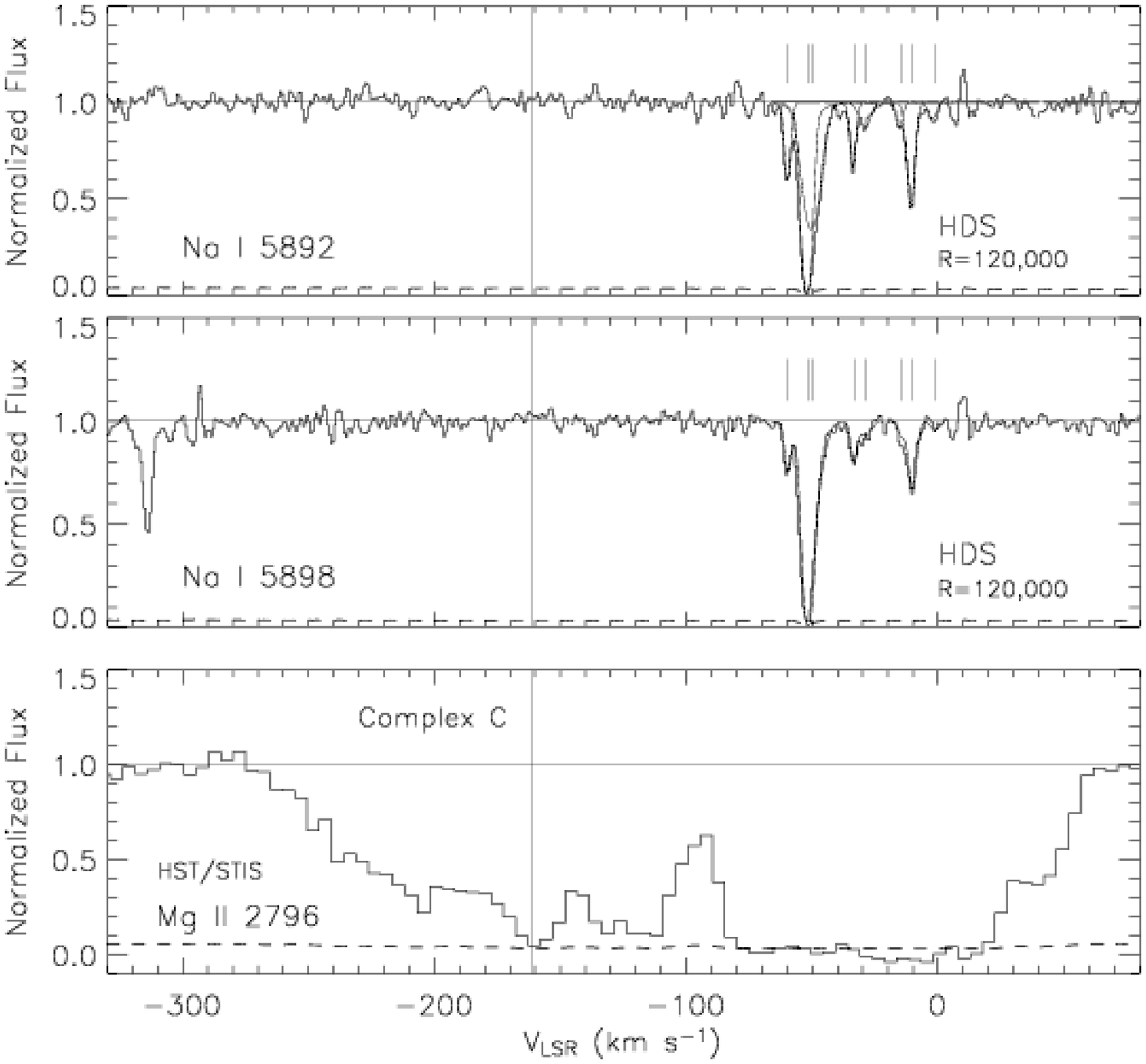}}
\protect
\caption{Galactic {\NaI} detections in the HDS spectrum of PG~$1634+706$. The superhigh resolution spectrum allowed measurements of $b$-value $\sim 1$ km s$^{-1}$. Superimposed in thin lines are the various Voigt Profile components contributing to the absorption. The vertical ticks mark the center of each component. The thin line in the {\NaI}~5898 panel shows the cumulative Voigt Profile by combining the various components. The fit results are listed in Table~\ref{tab:tab5}. The {\MgII} absorption from the HVC Complex C along the line of sight to the quasar is also shown. To guide the eye, the expected location of the {\NaI} doublet absorption from the HVC is marked by the vertical line.}
\label{fig:16}
\end{figure*}

In the HDS spectrum, Galactic {\NaI} is detected, as shown in
Figure~\ref{fig:16}.  Voigt profile fitting parameters are listed in
Table~\ref{tab:tab5}.  For the narrowest component, at $v_{LSR}=-32$~{\kms} {\footnote{We adopt the ``standard'' definition of the local standard of rest \citep{Kerr86}, in which the Sun is moving in the direction $\alpha=18h$, $\delta=+30^{o}$ (epoch $1900$) at $20$~{\kms}. With this convention, the conversion to heliocentric velocity is: $v_{helio} = v_{LSR}+14.770$~{\kms}.}}
we measure $b=0.9\pm0.2$~{\kms}. This shows that such narrow lines as
$b\sim1$~{\kms} can be measured at $R=120,000$. At the same time, we also note that the hyperfine splitting in {\NaI} lines are likely to affect our measurement of such a narrow line width. The difference in rest-frame wavelength from the basic hyperfine splitting of the D1 and D2 lines has been measured to be $0.0213$~{\AA} and $0.0198$~{\AA}, respectively \citep{morton03}. This corresponds to shifts in velocity of $1.0827$~{\kms} and $1.0075$~{\kms}, respectively. As these values are very close to our measurement of $b \sim 1$~{\kms}, we expect the true-$b$ to be lower than this measured value. However, observing this effect by resolving the line would require spectral resolutions that are significantly larger than our HDS measurement at $R=120,000$ (as discussed in \S~\ref{sec:2.2}.)

The PG~$1634+706$ line of sight passes through the well-studied high velocity
cloud complex C.  Upon inspection of a {\it HST}/STIS E230M spectrum, we
find absorption detected in {\MgII}, {\MgI}, and {\FeII} at $v_{LSR} \sim -115$
and $-155$~{\kms}.  We do not detect {\NaI}$\lambda\lambda$5892,5898
at these velocities in our $R=120,000$ spectrum to a $3\sigma$ limit of
$0.004$~{\AA} pixel$^{-1}$.  Recently, $21$cm {\HI}
observations have found that high velocity, low column density
hydrogen clumps have T~$< 900$~K \citep{hoffman04,richter05} and that
narrow {\CaII} lines are detected in most sightlines
\citep{richter05}.  Unfortunately, {\CaII}$\lambda\lambda$3935,3970 is
not covered in either our $R=45,000$ or $R=120,000$ spectrum of PG~$1634+706$.
However, at $R=45,000$, it would not be possible to test 
if the {\CaII} lines are as narrow as would be suggested by the low
temperatures from {\HI}.  Thus the study of HVCs is an application for
which super-high resolution will be particularly valuable.  This
applies to HVCs around galaxies at high redshift as well as to Milky
Way HVCs.

\section{CONCLUSION}
\label{sec:4}

The ability to routinely obtain quasar spectra with resolutions of
$R>100,000$ is not likely to provide as large a change in the field of
quasar absorption lines as the previous advance to $R=45,000$, made
possible with $8$--meter class telescopes.  However, the ability to
probe structure on scales of $\sim1$~{\kms} will certainly be valuable
in select applications and in detailed studies of particular
absorbers.  Examples of such applications include probing the physical
conditions of cold regions in the ISM of galaxies, particularly in
DLAs and other strong {\MgII} absorbers.  Also, it appears that high
velocity clouds, both around the Milky Way and at high redshift, may
have a cold phase. Finally, surveys extending to small equivalent
widths will be aided by the enhanced sensitivity of higher resolution
observations.

Our comparison of $R=120,000$ Subaru/HDS observations to $R=45,000$
Keck/HIRES observations of various systems toward PG~$1634+706$ show
that at the higher resolution, more accurate Doppler parameter
measurements can be made, and that additional Voigt profile components
are resolved.  However, for the particular {\MgII} absorbers along the
PG~$1634+706$ sightline no extremely narrow individual components were
measured.  It was only for the Galactic {\NaI} line in this line of
sight that $b<1$~{\kms} was measured.  Because such narrow components
are expected in a variety of environments, it would be valuable to
push the limits of 8-meter class telescopes in order to detect them in
a variety of types of systems toward the brightest quasars.

We are grateful to Chris Churchill for giving us permission to use the Keck/HIRES spectrum for this work.  We also wish to thank the referee, John Black, for several insightful comments that have increased the scope of this paper.



\begin{figure*}
\figurenum{10a}
\epsscale{0.8}
\plotone{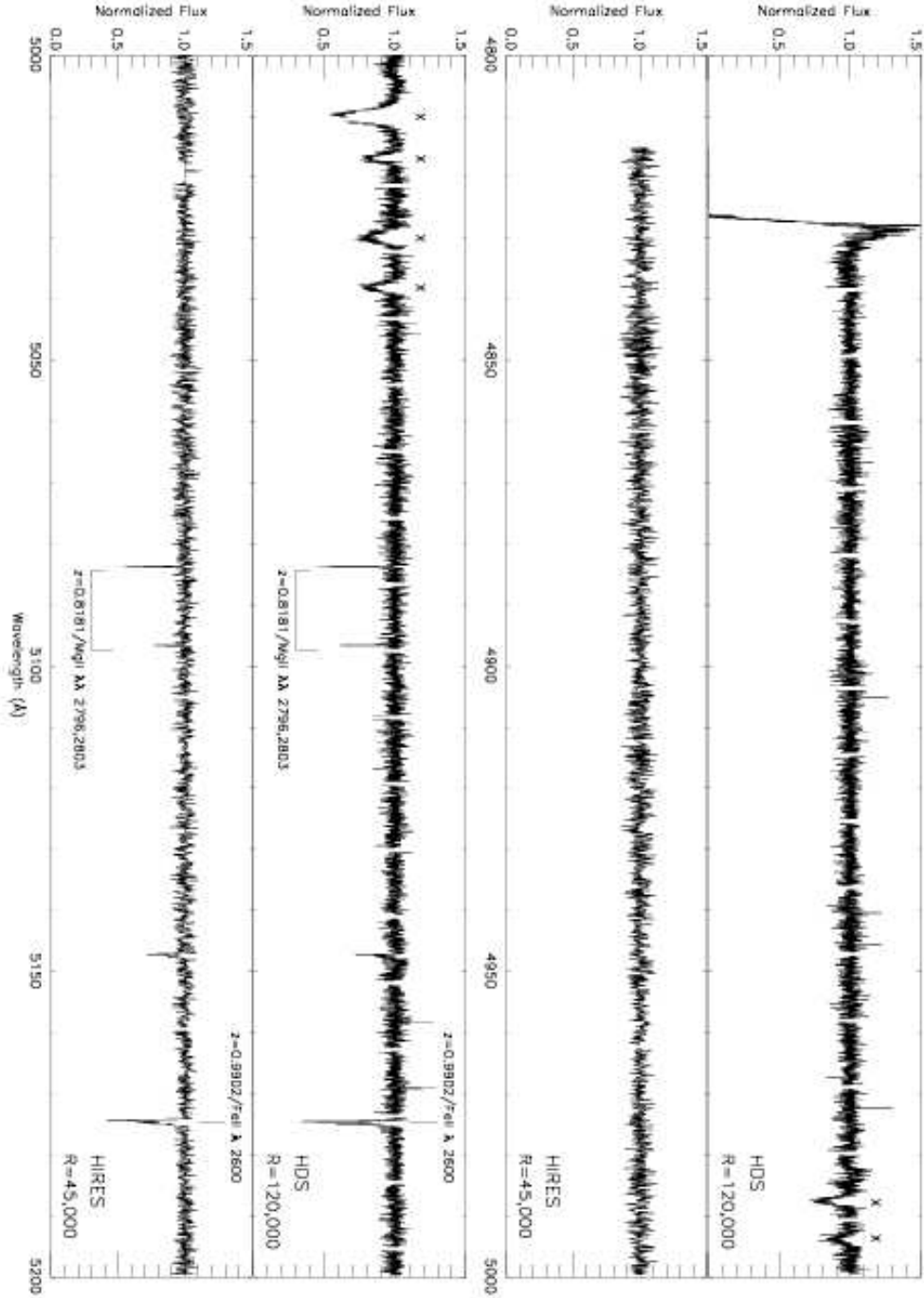}
\protect
\caption{Comparison of PG~$1634+706$ observations using $R=120,000$ Subaru/HDS and $R=45,000$ Keck/HIRES. The details of the observation are listed in Table~\ref{tab:tab1}. The metal lines associated with the absorption systems discussed in this paper are identified and labelled in both spectrum. The features in the HDS spectrum marked with 'x' are unreal and were produced by bad regions in the CCD.}
\label{fig:10a}
\end{figure*}

\begin{figure*}
\figurenum{10b}
\epsscale{0.8}
\plotone{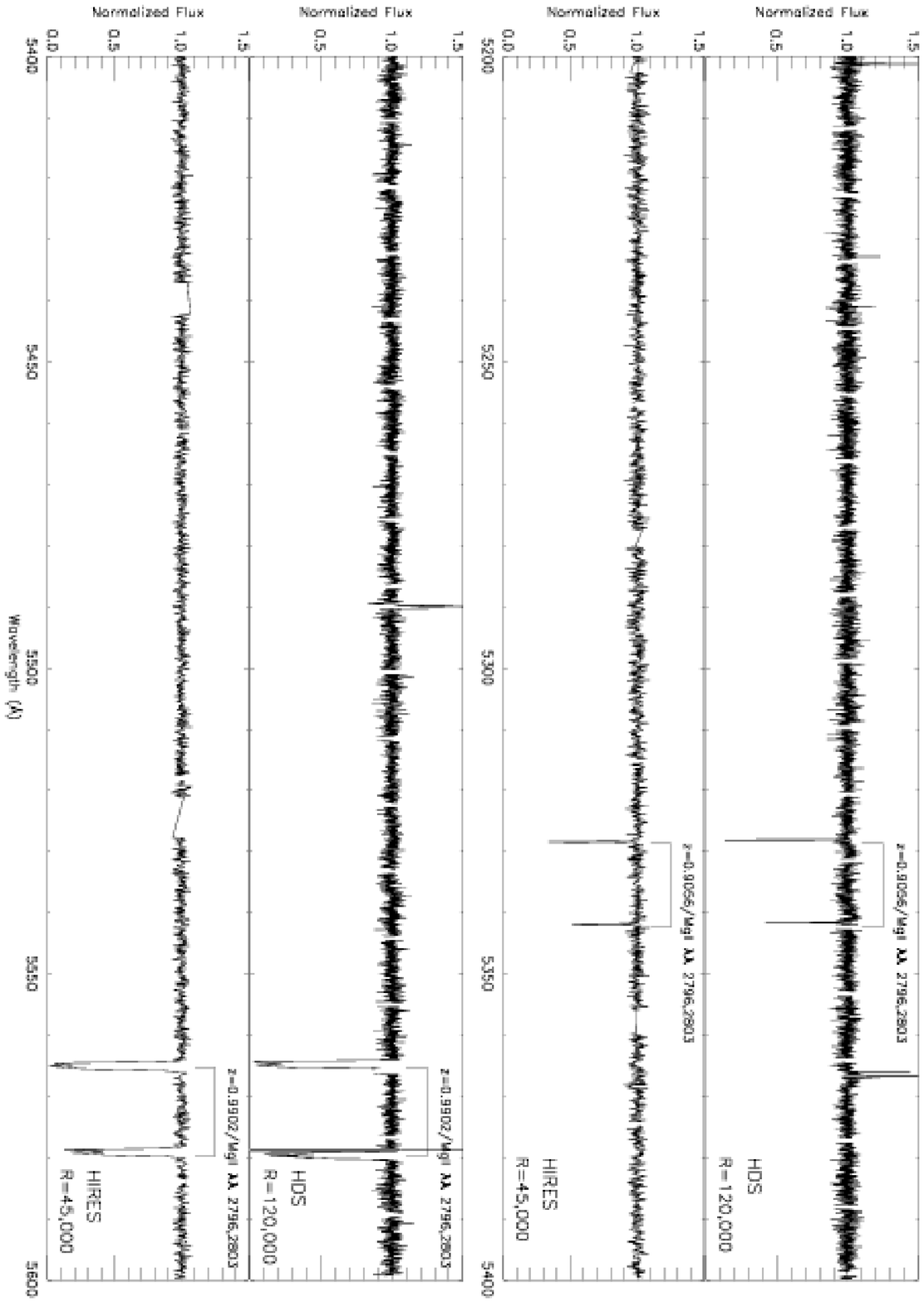}
\protect
\caption{Contd. Figure~\ref{fig:10a}}
\label{fig:10b}
\end{figure*}

\begin{figure*}
\figurenum{10c}
\epsscale{0.8}
\plotone{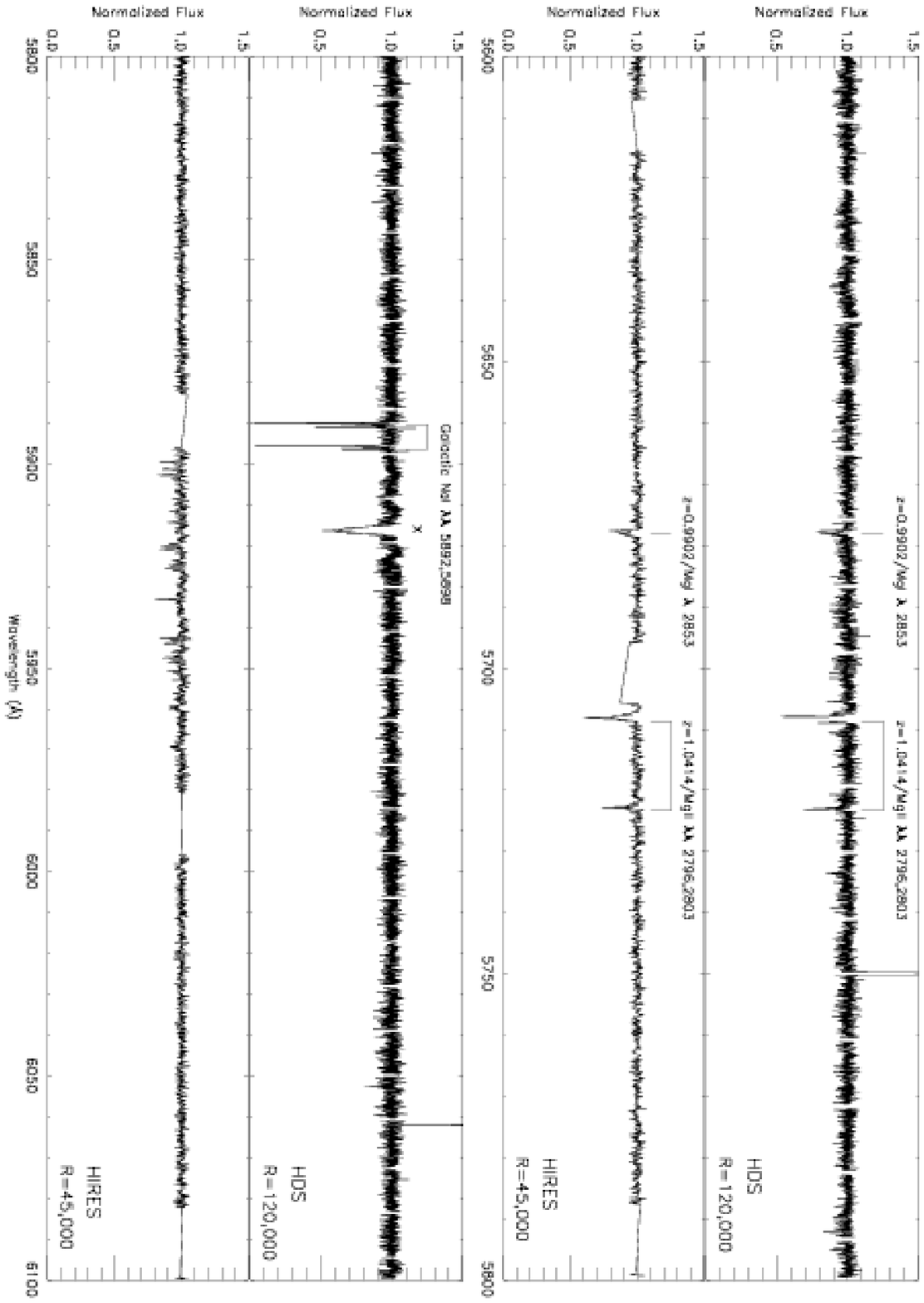}
\protect
\caption{Contd. Figure~\ref{fig:10a}}
\label{fig:10c}
\end{figure*}


\newpage

\clearpage
\begin{deluxetable}{lccccccc}
\tablenum{1}
\tabletypesize{\footnotesize}
\tablewidth{0pt}
\tablecaption{\textsc{Results from simulations}}
\tablehead{
\multicolumn{1}{l}{$R/1000$} &
\multicolumn{1}{l}{$S/N$ ratio} &
\multicolumn{3}{|c|}{\textbf{$b({\Mgi}) = 2.0$~{\kms}}} &
\multicolumn{3}{c}{\textbf{$b({\Mgi}) = 1.0$~{\kms}}}\\
\hline
\colhead{} &
\colhead{pixel$^{-1}$} &
\colhead{$b({\Mgi})^*$} &
\colhead{$\log$ $N({\Mgi})^*$} &
\colhead{\% Models $^{\dag}$} &
\colhead{$b({\Mgi})^*$} &
\colhead{$\log$ $N({\Mgi})^*$} &
\colhead{\% Models $^{\dag}$}
}
\startdata
{$120$} & {$35$} & $2.02_{-0.14}^{+0.15}$ & $12.0_{-0.02}^{+0.02}$ & $99.99$ & $1.07_{-0.16}^{+0.16}$ & $11.97_{-0.04}^{+0.06}$ & $90.21$ \\
\\
{$45$} & {$80$} & $2.15_{-0.48}^{+0.43}$ & $11.99_{-0.03}^{+0.04}$ & $94.33$ & $1.82_{-0.33}^{+0.44}$ & $11.84_{-0.03}^{+0.04}$ & $44.00$ \\
\\
\hline
\\
{$120$} & {$55$} & $2.02_{-0.09}^{+0.10}$ & $12.0_{-0.02}^{+0.01}$ & $99.98$ & $1.05_{-0.13}^{+0.11}$ & $11.97_{-0.03}^{+0.05}$ & $97.80$ \\
\\
{$45$} & {$128$} & $2.17_{-0.32}^{+0.26}$ & $11.98_{-0.01}^{+0.03}$ & $98.44$ & $1.56_{-0.27}^{+0.35}$ & $11.85_{-0.02}^{+0.04}$ & $51.86$ \\
\\
\hline
\\
{$120$} & {$70$} & $2.02_{-0.07}^{+0.08}$ & $12.0_{-0.01}^{+0.01}$ & $99.98$ & $1.05_{-0.10}^{+0.09}$ & $11.97_{-0.02}^{+0.04}$ & $98.98$ \\
\\
{$45$} & {$160$} & $2.17_{-0.25}^{+0.21}$ & $11.98_{-0.01}^{+0.02}$ & $98.03$ & $1.55_{-0.28}^{+0.33}$ & $11.86_{-0.03}^{+0.03}$ & $51.98$ \\
\\
\enddata
\label{tab:tab1}
\tablecomments{$^*$ Median $b$ and $N$ values. \\
$^{\dag}$ - Out of 10,000 realizations, the percentage of Voigt Profile fit models that gave a reliable measurement ($\sigma(b)<b$).}
\end{deluxetable}

\clearpage
\begin{deluxetable}{lcccccl}
\tablenum{2}
\tablecaption{PG~$1634+706$ \textsc {Observations}; $z_{\textrm{\scriptsize QSO}}$=$1.337$, $m_V$=14.9.} 
\tablehead{
\colhead{Instrument} & 
\colhead{Resolution} &
\colhead{${\triangle}v$} & 
\colhead{Date(s) of} & 
\colhead{T$_{exp}$} & 
\colhead{Coverage} &
\colhead{PI} \\
\colhead{} &
\colhead{$\lambda$/${\triangle}\lambda$} &
\colhead{km s$^{-1}$} &
\colhead{Observation} &
\colhead{seconds} &
\colhead{$\lambda$ {\AA}} &
\colhead{}
} 
\startdata
{Keck/HIRES} & $45,000$ & $6.6$ & $1994$ Jul $04$ & $2700$ & $3723-6185$ & Churchill \\
{ }	     & $45,000$ & $6.6$ & $1994$ Jul $05$ & $5400$ & $3723-6185$ & Churchill \\

{Subaru/HDS} & $120,000$ & $2.5$ & $2005$ Jun $06$ & $3600$ & $4830-7560$ & Misawa \\
\enddata
\label{tab:tab2}
\end{deluxetable}

\clearpage
\begin{deluxetable}{lrrr}
\tablenum{3}
\tablewidth{0pt}
\tablecaption{\textsc{Voigt Profile Fit Results for $z=0.9902$ Strong {\MgII} System}} 
\tablehead{ 
\colhead{Instrument} &
\colhead{$v$} &
\colhead{$\log$~$N$} & 
\colhead{$b$} \\
\colhead{} &
\colhead{km s$^{-1}$} &
\colhead{atoms cm$^{-2}$} &
\colhead{km s$^{-1}$}
} 
\startdata
\multicolumn{4}{c}{{\MgI} \textsf{Fit Results}}\\
\\
{HIRES} & $-22.6$ & $11.35~{\pm}~0.04$ & $6.52~{\pm}~0.71$ \\
{ } & $5.2$ & $11.37~{\pm}~0.06$ & $13.52~{\pm}~2.44$ \\
{ } & $25.0$ & $10.65~{\pm}~0.18$ & $3.2~{\pm}~2.52$ \\
\\
{HDS} & $-20.1$ & $11.26~{\pm}~0.03$ & $5.87~{\pm}~0.46$ \\
{ } & $7.7$ & $11.28~{\pm}~0.04$ & $10.55~{\pm}~1.28$ \\
{ } & $24.5$ & $10.51~{\pm}~0.10$ & $5.87~{\pm}~0.94$ \\
\\
\hline
\hline
\\
\multicolumn{4}{c}{{\MgII} \textsf{Fit Results}}\\
\\
{HIRES} & $-22.1$ & $13.04~{\pm}~0.01$ & $7.72~{\pm}~0.09$ \\
{ } & $1.1$ & $12.82~{\pm}~0.01$ & $4.11~{\pm}~0.12$ \\
{ } & $8.3$ & $13.30~{\pm}~0.01$ & $22.29~{\pm}~0.21$ \\
\\
{HDS} & $-20.2$ & $13.05~{\pm}~0.01$ & $7.57 {\pm}~0.15$ \\
{ } & $-8.2$ & $12.44~{\pm}~0.06$ & $4.46~{\pm}~0.48$ \\
{ } & $3.8$ & $12.90~{\pm}~0.22$ & $6.50~{\pm}~1.00$ \\
{ } & $16.8$ & $12.93~{\pm}~0.34$ & $11.03~{\pm}~8.00$ \\
{ } & $30.1$ & $12.34~{\pm}~0.55$ & $6.11~{\pm}~1.91$ \\
{ } & $45.4$ & $11.34~{\pm}~0.09$ & $3.37~{\pm}~0.69$ \\
\\
\hline
\enddata
\label{tab:tab3}
\end{deluxetable}

\clearpage
\begin{deluxetable}{lrrr}
\tablenum{4}
\tablewidth{0pt}
\tablecaption{\textsc{Voigt Profile Fit Results for Weak {\MgII} Systems}} 
\tablehead{ 
\colhead{Instrument} &
\colhead{$v$} &
\colhead{$\log$~[$N({\Mgi})$]} & 
\colhead{$b({\Mgi})$} \\
\colhead{} &
\colhead{km s$^{-1}$} &
\colhead{atoms cm$^{-2}$} &
\colhead{km s$^{-1}$}
} 
\startdata
\multicolumn{4}{c}{$z = 0.8181$ single--cloud system} \\
\hline
\\
{HIRES} & $0$ & $12.04~{\pm}~0.03$ & $2.14~{\pm}~0.40$ \\
{HDS} & $0$ & $11.97~{\pm}~0.02$ & $2.61~{\pm}~0.13$ \\
\\
\hline
\multicolumn{4}{c}{$z = 0.9056$ single--cloud system} \\
\hline
\\
{HIRES} & $0$ & $12.47~{\pm}~0.01$ & $2.77~{\pm}~0.12$ \\
{HDS} & $0$ & $12.42~{\pm}~0.01$ & $3.14~{\pm}~0.07$ \\
\\
\hline
\multicolumn{4}{c}{$z = 1.0414$ multiple--cloud system} \\
\hline
\\
{HIRES} & $-38.2$ & $11.49~{\pm}~0.08$ & $6.53~{\pm}~1.62$ \\
{ } & $-15.2$ & $11.90~{\pm}~0.06$ & $10.81~{\pm}~2.03$ \\
{ } & $0$ & $12.02~{\pm}~0.03$ & $3.12~{\pm}~0.38$ \\
{ } & $15.7$ & $11.68~{\pm}~0.04$ & $5.95~{\pm}~0.83$ \\
\\
{HDS} & $-36.4$ & $11.17~{\pm}~0.23$ & $3.98~{\pm}~2.78$ \\
{ } & $-29.6$ & $10.98~{\pm}~0.27$ & $2.23~{\pm}~1.46$ \\
{ } & $-17.3$ & $11.38~{\pm}~0.14$ & $5.19~{\pm}~1.99$ \\
{ } & $-10.1$ & $11.05~{\pm}~0.25$ & $2.35~{\pm}~1.08$ \\
{ } & $1.9$ & $12.02~{\pm}~0.01$ & $3.82~{\pm}~0.13$ \\
{ } & $18.0$ & $11.48~{\pm}~0.04$ & $2.86~{\pm}~0.34$ \\
\\
\enddata
\label{tab:tab4}
\end{deluxetable}

\clearpage
\begin{deluxetable}{lrr}
\tablenum{5}
\tablewidth{0pt}
\tablecaption{\textsc{Voigt Profile Measurement of Galactic {\NaI} absorption in the HDS spectrum of PG~$1634+706$}}
\tablehead{ 
\colhead{$v_{LSR}$} &
\colhead{$\log$~[$N$(Na)]} & 
\colhead{$b$~(Na)} \\
\colhead{km s$^{-1}$} &
\colhead{atoms cm$^{-2}$} &
\colhead{km s$^{-1}$}
} 
\startdata
$-1.1$ & $10.56~{\pm}~0.10$ & $1.68~{\pm}~0.82$ \\
$-10.0$ & $11.51~{\pm}~0.02$ & $1.63~{\pm}~0.16$ \\
$-14.4$ & $10.67~{\pm}~0.10$ & $0.89~{\pm}~0.88$ \\
$-28.6$ & $10.90~{\pm}~0.11$ & $2.80~{\pm}~1.02$ \\
$-33.2$ & $11.14~{\pm}~0.05$ & $0.90~{\pm}~0.33$ \\
$-49.9$ & $11.96~{\pm}~0.34$ & $4.64~{\pm}~0.28$ \\
$-51.7$ & $14.21~{\pm}~1.63$ & $1.02~{\pm}~0.35$ \\
$-59.6$ & $11.26~{\pm}~0.05$ & $0.86~{\pm}~0.21$ \\
\enddata
\label{tab:tab5}
\end{deluxetable}


\begin{thebibliography}{XXX}

\bibitem[Bergeron \& Boiss{\'e}(1991)]{bb91}
Bergeron, J., \& Boiss{\'e}, P. 1991, AA, 243,344

\bibitem[Beverini {\etal}(1990)]{beverini}
Beverini, N., Maccioni, E., Pereira, D., Strumia, F., Vissani, G., \& Wang, Y.-Z. 1990, Optics Communications, 77, 299

\bibitem[Caulet (1989)]{caulet}
Caulet, A. 1989, ApJ, 340, 90

\bibitem[Chand {\etal}(2006)]{chand06}
Chand, H., Srianand, R., Petitjean, P., Aracil, B., Quast, R., \& Reimers, D. 2006, AA, 451, 45

\bibitem[Charlton \& Churchill(1998)]{charcwc98}
Charlton, J. C., \& Churchill, C. W. 1998, ApJ, 499, 181

\bibitem[Charlton {\etal}(2003)]{weak1634} 
Charlton, J. C., Ding, J., Zonak, S. G., Churchill, C. W., Bond, N. A., \& Rigby, J. R. 2003, ApJ, 589, 111

\bibitem[Charlton {\etal}(2000)]{jane00}
Charlton, J. C., Mellon, R. R., Rigby, J. R., \& Churchill, C. W. 2000, ApJ, 545, 635

\bibitem[Churchill (1997)]{cwcthesis}
Churchill, C. W. 1997, Ph.D. Thesis, University of Santa Cruz

\bibitem[Churchill {\etal}(2000)]{archive2}
Churchill, C. W., Mellon, R. R., Charlton, J. C., Jannuzi, B. T., Kirhakos, S., Steidel, C. C., \& Schneider, D. 2000, ApJ, 543, 577

\bibitem[Churchill {\etal}(1999)]{weak1}
Churchill, C. W., Rigby, J. R., Charlton, J. C., \& Vogt, S. S. 1999, ApJS, 120, 51 

\bibitem[Churchill {\etal}(2001)]{cwcvogt01}
Churchill, C. W., \& Vogt, S. S. 2001, AJ, 122, 679

\bibitem[Churchill {\etal}(2003)]{cwcvogtchar03}
Churchill, C. W., Vogt, S. S., \& Charlton, J. C. 2003, AJ, 125, 98

\bibitem[Cowie et al.(1995)]{cowie95} Cowie, L.~L., Songaila, 
A., Kim, T.-S., \& Hu, E.~M.\ 1995, \aj, 109, 1522 

\bibitem[Dav{\'e} {\etal}(1996)]{dave96}
Dav{\'e}, R., Hernquist, L., Katz, N., Weinberg, D., \& Churchill, C. W. 1996, AAS, 188.2308D

\bibitem[Dekker {\etal}(2000)]{uves}
Dekker, H., {\etal}2000, SPIE, 4008, 534

\bibitem[Ding {\etal}(2003)]{z99}
Ding, J., Charlton, J. C., Bond, N. A., Zonak, S. G., \& Churchill, C. W. 2003, ApJ, 587, 551

\bibitem[Ding {\etal}(2005)]{ding05}
Ding, J., Charlton, J. C., \& Churchill, C. W. 2005, ApJ, 621, 615

\bibitem[Elmegreen (1997)]{elmegreen}
Elmegreen, B. G. 1997, ApJ, 477, 196

\bibitem[Fox {\etal}(2005)]{fox05}
Fox, A. J., Wakker, B. P., Savage, B. D., Tripp, T. M., Sembach, K. R., \& Bland-Hawthorn, J. 2005, ApJ, 630, 332

\bibitem[Hoffman {\etal}(2004)]{hoffman04}
Hoffman, G. L., Salpeter, E. E., \& Hirani, A. 2004, AJ, 128, 2932

\bibitem[Jones {\etal}(2006)]{jones06}
Jones, T., Narayanan, A., Mshar, A., \& Charlton, J. C. 2006, {\it in preparation}

\bibitem[Kerr \& Lynden-Bell(1986)]{Kerr86}
Kerr, F. J., \& Lynden-Bell, D. 1986, MNRAS, 221, 1023

\bibitem[Lane {\etal}(2000)]{lane00}
Lane, W. M., Briggs, F. H., \& Smette, A. 2000, ApJ, 532, 146

\bibitem[Lanzetta \& Bowen(1992)]{lanzettabowen}
Lanzetta, K. M., \& Bowen, D. V. 1992, ApJ, 391, 48

\bibitem[Lanzetta {\etal}(1987)]{lanzetta87}
Lanzetta, K. M., Turnshek, D. A., \& Wolfe, A. M. 1987, ApJ, 322, 739

\bibitem[Lynch {\etal}(2006)]{lynch06}
Lynch, R. S., Charlton, J. C., Kim, T. S. 2006, ApJ, 640, 81

\bibitem[Masiero {\etal}(2005)]{masiero05}
Masiero, J. R., Charlton, J. C., Ding, J., Churchill, C. W., Kacprzak, G. 2005, ApJ, 623, 57

\bibitem[Misawa et al.(2002)]{toru02} Misawa, T., Tytler, D., 
Iye, M., Storrie-Lombardi, L.~J., Suzuki, N., \& Wolfe, A.~M.\ 2002, \aj, 
123, 1847 

\bibitem[Morton (2003)]{morton03}
Morton, D. C. 2003, ApJS, 149, 205

\bibitem[Narayanan {\etal}(2005)]{anand05}
Narayanan, A., Charlton, J. C., Masiero, J. R., \& Lynch, R. 2005, ApJ, 632, 92

\bibitem[Noguchi {\etal}(1999)]{hds}
Noguchi, K., {\etal}1999, PASJ, 54, 855

\bibitem[Petitjean {\etal}(2000)]{petitjean}
Petitjean, P., Srianand, R., \& Ledoux, C. 2000, AA, 364, 26

\bibitem[Points {\etal}(2004)]{points}
Points, S. D., Lauroesch, J. T., \& Meyer, D. M. 2004, PASP, 116, 801

\bibitem[Richter {\etal}(2005)]{richter05}
Richter, P., Westmeier, T., \& Br{\.u}ns, C. 2005, AA, 442L, 49

\bibitem[Rigby {\etal}(2002)]{weak2}
Rigby, J. R., Charlton, J. C., \& Churchill, C. W. 2002, ApJ, 565, 743

\bibitem[Sargent et al.(1988)]{sbs88} 
Sargent, W.~L.~W., Boksenberg, A., \& Steidel, C.~C.\ 1988, \apjs, 68, 539 
 
\bibitem[Sembach (2002)]{sembach}
Sembach, K. R. 2002, ASPC, 254, 283 (astro-ph/0108088)

\bibitem[Srianand {\etal}(2005)]{srianand05}
Srianand, R., Petitjean, P., Ledoux, C., Ferland, G., \& Shaw, G. 2005, MNRAS, 362, 549

\bibitem[Steidel {\etal}(1997)]{steidel97}
Steidel, C. C., Dickinson, M., Meyer, D. M., Adelberger, K. L., \& Sembach, K. R. 1997, ApJ, 480, 568

\bibitem[Tappe \& Black(2004)]{tappeblack}
Tappe, A., \& Black, J. H 2004, AA, 423, 943

\bibitem[Vogt {\etal}(1994)]{hires}
Vogt, S. S., {\etal}1994, Proc.SPIE, 2198, 362

\bibitem[Wolfe {\etal}(2003)]{wolfe03}
Wolfe, A, M., Prochaska, J. X., \& Gawiser, E. 2003, ApJ, 593, 215

\bibitem[Zonak {\etal}(2004)]{zonak}
Zonak, S. G., Charlton, J. C., Ding, J., \& Churchill, C. W. 2004, ApJ, 606, 196

\end{thebibliography}
\end{document}